\documentclass[12pt,preprint]{aastex}
\usepackage{emulateapj5, apjfonts}
\usepackage{epsfig}

\newcommand{\msun}{M$_{\odot}$}
\newcommand{\ergl}{ergs~s$^{-1}$}

\newcommand{\ergcms}{ergs~cm$^{-2}$~s$^{-1}$}
\newcommand{\fcgs}{${\rm erg}\, {\rm cm}^{-2}\, {\rm s}^{-1}$}
\newcommand{\lcgs}{${\rm erg}\, {\rm s}^{-1}$}

\newcommand{\chisq}{\chi^2}
\newcommand{\CXO}{{\sl CXO}}
\newcommand{\ROSAT}{{\sl ROSAT}}
\newcommand{\HST}{{\sl HST}}

\newcommand{\psr}{PSR B1821$-$24}
\newcommand{\etal}{et~al.}

\slugcomment{\apj \, accepted for publication}
\shorttitle{{\sl Chandra} observations of the globular cluster M28}
\shortauthors{}

\begin{document}

\title{{\sl Chandra} X-Ray Observatory observations of the globular cluster 
M28\\[1ex] and its millisecond pulsar \psr}

\author{
Werner Becker\altaffilmark{1},
Douglas A. Swartz\altaffilmark{2},
George G. Pavlov\altaffilmark{6},
Ronald F. Elsner\altaffilmark{3},
Jonathan Grindlay\altaffilmark{4},
Roberto Mignani\altaffilmark{5},
Allyn F. Tennant\altaffilmark{3},
Don Backer\altaffilmark{7},\\
Luigi Pulone\altaffilmark{8},
Vincenzo Testa\altaffilmark{8}\\ and\\ Martin C. Weisskopf\altaffilmark{3}\vspace{0.2cm} }

\altaffiltext{1}
{Max Planck Institut f\"ur Extraterrestrische Physik, 85741 Garching bei
M\"unchen, Germany}
\altaffiltext{2}
{USRA, Space Science Department, NASA Marshall Space Flight Center, SD50,
Huntsville, AL 35812}
\altaffiltext{3}
{Space Science Department, NASA Marshall Space Flight Center, SD50, 
Huntsville, 
AL 35812}
\altaffiltext{4}
{Harvard-Smithsonian Center for Astrophysics, 60 Garden street, Cambridge, MA 
02138}
\altaffiltext{5}
{European Southern Observatory, 85740 Garching bei M\"unchen, Germany}
\altaffiltext{6}
{The Pennsylvania State University, 525 Davey Lab, University Park, PA 16802}
\altaffiltext{7}
{The University of California, 415 Campbell Hall, Berkely, CA 94720-3411}
\altaffiltext{8}
{INAF - Osservatorio Astronomico di Roma, Via Frascati 33  00040 Monte  
Porzio Catone (Italy)}

\begin{abstract}

We report here the results of the first {\sl Chandra} X-Ray Observatory
observations of the globular  cluster M28 (NGC 6626). 
We detect 46 X-ray sources of which 12 lie within one core radius of the 
center.
We  show  that the apparently extended X-ray core emission seen with the ROSAT
HRI is due to the superposition of multiple discrete sources for which we
determine the X-ray luminosity function down to a limit of about
$6\times10^{30}$ erg s$^{-1}$. 
We measure the radial distribution of the X-ray sources and fit it to a King
profile finding a core radius of $r_{c,x} \approx 11\arcsec$.
We measure for the first time the unconfused phase-averaged X-ray spectrum of
the 3.05-ms pulsar B1821$-$24 and find it is best described by a power law 
with
photon index $\Gamma\simeq 1.2$.
We find marginal evidence of an emission line centered at 3.3 keV in the 
pulsar
spectrum, which could be interpreted as cyclotron emission from a corona above
the pulsar's polar cap if the the magnetic field is strongly different from a
centered dipole. 
The unabsorbed pulsar
flux in the 0.5--8.0 keV band is $\approx 3.5\times 10^{-13}\,{\rm ergs\,\,
s}^{-1}\,{\rm cm}^{-2}$.
We present spectral analyses of the 5 brightest unidentified sources. Based on
the spectral parameters
of the brightest of these sources, we suggest that it is a transiently 
accreting
neutron star 
in a low-mass X-ray binary, in quiescence. Fitting its spectrum with a 
hydrogen
neutron star 
atmosphere model yields the effective temperature $T_{\rm eff}^\infty =
90^{+30}_{-10}$ eV 
and the radius $R_{\rm NS}^\infty = 14.5^{+6.9}_{-3.8}$ km. In addition to the
resolved sources, 
we detect fainter, unresolved X-ray emission from the central core. Using the
{\sl Chandra}-derived 
positions, we also report on the result of searching archival Hubble Space
Telescope data for 
possible optical counterparts.

\end{abstract}


\section{Introduction}

Since the {\sl Einstein} era it has been clear that globular clusters contain
various populations of
X-ray sources of very different luminosities (Hertz \& Grindlay 1983).
The stronger sources ($L_{x} \approx 10^{36}-10^{38}$ \lcgs) were seen to  
exhibit X-ray bursts which led to their identification as low-mass X-ray
binaries (LMXBs). The nature of the weaker sources, with  
$L_x \le 3\!\times \! 10^{34}$ \lcgs,  however was more open to discussion 
(e.g., Cool et  al.\ 1993; Johnston \& Verbunt 1996). Although many weak 
X-ray sources were detected in globulars by ROSAT (Johnston \& Verbunt 1996; 
Verbunt 2001), their identification has been difficult due to low photon 
statistics and strong source confusion in the crowded globular cluster fields, 
except for a few cases (Cool et al.\ 1995 and Grindlay et al.\ 1995).
The application of the {\sl Chandra} X-Ray Observatory (\CXO) sub-arcsecond 
angular resolution to the study of globular cluster X-ray sources leads one 
to anticipate progress in our understanding.

Advances with the \CXO\ have already included observations of NGC 6752 
(Pooley et al.\ 2002a), 6440 (Pooley et al.\ 2002b), 6397 (Grindlay et  
al.\ 2001) and 5139 (Rutledge et al.\ 2002a) which detected more  
low-luminosity X-ray sources than in all ROSAT observations of 55 globular  
clusters combined. Of particular interest are the results obtained from 
\CXO\ observations of 47~Tuc = NGC 104. Grindlay et al.~(2001) reported 
the detection of 108 sources within a region  corresponding to about 5 
times the 47~Tuc core radius. Sixteen of the soft/faint sources were found 
to be coincident with radio-detected millisecond pulsars (MSPs), and 
Grindlay et al.\ (2001, 2002) concluded that more than 50 percent of all 
the unidentified sources in 47~Tuc are MSPs. This conclusion is in concert 
with theoretical estimates on the formation scenarios of short-period 
(binary) pulsars in globular clusters (Rasio, Pfahl \& Rappaport 2000).

The globular cluster M28 = NGC 6626 lies close to the galactic plane
($b=-5.58\degr$) and close to the galactic center ($l=7.8\degr$) (Harris 
1996).
Distance estimates for M28 range from 5.1 kpc (Rees \& Cudworth 1991) to 5.7 
kpc (Harris 1996). In this paper, we use 5.5 kpc as a reference distance.

M28 is a relatively compact cluster with a core radius of 0\farcm24,
corresponding 
to $r_c\sim 0.4$ pc, and a half-mass radius of 1\farcm56, 
corresponding to $\sim$2.6 pc (Harris 1996).  The values of these radii in
parsec 
are smaller than those for the better studied 47~Tuc. Thus, although M28's
central 
luminosity density, $\rho_0 = 10^{4.75}\,\, L_\odot\,\, {\rm pc}^{-3}$, is 
comparable to that of 47~Tuc ($10^{4.77}\,\, L_\odot\,\, {\rm pc}^{-3}$ ---
Harris 1996), 
the rate of two-body encounters in the core, $N_{\rm enc} \propto \rho_0^{1.5}
r_c^2$, 
is a factor of $2$ smaller, and thus fewer binaries are created and expected 
as
dim 
X-ray sources~\footnote{$N_{\rm enc}\propto \rho_0^2  r_c^3 v_{\rm disp}^{-1}$ 
(cf.\ Verbunt and Hut 1987) and $v_{\rm disp} \sim (GM/r_c)^{0.5} \propto 
\rho_0^{0.5} r_c$, where $M\propto \rho_0 r_c^3$ is the mass of the core of 
the
cluster (see also Verbunt 2002).}.

Davidge, Cote \& Harris (1996) noted that M28's position in the plane
near the galactic center and its relative compactness may indicate that it has 
been in the inner Galaxy for a long time. The authors suggest an age of 
$\sim$16 Gyr. This age, however, seems to be too high compared with the
recent results by Salaris \& Weiss (2002) and Testa et al.~(2001) from
which one estimates the age of M28 to be $11.4-11.7$ Gyr, consistent 
with M28's moderately low metallicity [Fe/H]$ = -1.45$.

The absorbing column towards M28 is $\sim$10 times larger than 
for  47~Tuc, with reddening  E$(B-V) = 0.43$ (Harris 1996) corresponding 
to a hydrogen column density $N_H\sim 2.4 \times 10^{21}$ cm$^{-2}$.

The first millisecond pulsar discovered in a globular cluster was \psr\ in M28
(Lyne et al.\ 1987). This solitary MSP has a rotation period of $P = 3.05$ ms
and
period derivative of $\dot{{P}}= 1.61\times 10^{-18}$ s s$^{-1}$.
This value for the period derivative is sufficiently large that the expected 
correction due to line-of-sight projection of acceleration in the  
cluster's gravitational potential (Phinney 1993) is $\la$10\% (assuming a  
$10''$ projected offset from the cluster center and a cluster core mass  
$\sim 10^4$~\msun; this assumed core mass is slightly higher than that  
derived from the values for central luminosity density and core radius  
given by Harris 1996). Therefore, the measured value reasonably accurately 
reflects the pulsar's intrinsic period derivative. The inferred pulsar 
parameters make it the youngest ($P/2\dot{P}=3.0\times 10^7$ yrs) and most 
powerful ($\dot{E}=2.24\times 10^{36} I_{45}$ erg s$^{-1}$) pulsar among 
all known MSPs. Here $I_{45}$ is the neutron star moment of inertia in 
units of $10^{45}$ g~cm$^{2}$. For the idealized magnetic dipole radiation 
model, the inferred perpendicular component of the magnetic dipole moment 
is $\mu\,\sin\alpha = 3.2\times10^{37} (I_{45} P \dot{P})^{1/2} = 
2.2\times10^{27} I_{45}^{1/2}$ G~cm$^3$, where $\mu$ is the magnetic 
dipole moment and $\alpha$ the angle between the rotation and magnetic 
dipole axes. Neglecting higher multipole contributions to the magnetic 
field and assuming the magnetic moment to be at the neutron star center, 
the inferred magnitude of the dipolar field at the magnetic pole is 
$B_p = 4.5\times 10^9\,  I_{45}^{1/2} R_{6}^{-3} (\sin\alpha)^{-1}$ G, 
where $R_{6}$ is the neutron star radius in units of $10^6$ cm.
Although these values for the dipole moment and polar dipole field are about 2 
orders of magnitude smaller than what is inferred for ordinary field pulsars 
in the galactic plane, they are the highest among all MSPs. The pulsar's high 
rotational-energy loss made \psr\  a prime candidate to be a rotationally
powered, 
non-thermal X-ray source. This was  confirmed by  X-ray observations performed 
with  different X-ray satellites, e.g., ROSAT (Danner et al.\ 1997 --- 
cf.\ Verbunt  2001), {\sl ASCA} (Saito  et al.\ 1997) and  {\sl  RXTE}\ 
(Rots  et  al.\  1998), which  detected  Crab-like pulsations.

While  the  detection of  strongly  pulsed  emission  would suggest a 
magnetospheric origin of the  X-ray emission, a clear characterization 
of  the \psr\ spectrum has been hampered so far by the crowding of sources 
in the region. Indeed, observation of M28  with the ROSAT HRI made it clear 
that all spectral data obtained from \psr, especially in the ``soft''  bands 
$\le 10$~keV, suffer from  spectral contamination from nearby sources.
Stretching the angular resolution of the HRI to its limit, four X-ray
sources within 2\arcmin\ of the cluster center were discovered, including  
the two barely resolved sources RX J1824.5$-$2452E and RX J1824$-$2452P 
located  
within 15\arcsec\ of the center of M28, with the latter being identified from 
the timing as the counterpart of the millisecond pulsar B1821$-$42.

Danner et al.\ (1997) put forth the suggestions that the $\sim 10'' \times  
10\arcsec$ extended region of emission associated with RX J1824.5--2452E 
was due to either a synchrotron nebula powered by the pulsar or a number 
of faint LMXBs. The latter interpretation was favored as measurements with 
the ROSAT PSPC exhibited flux variations by as much as a factor of 3 between 
observations performed in 1991 and 1995 (see also Becker \& Tr\"umper 1999).
Further evidence was provided by Verbunt (2001) who reanalyzed the ROSAT HRI 
data and resolved RX J1824.5$-$2452E into at least 2 sources. Time variations 
of the sources in M28 are also compatible with the result of Gotthelf \&  
Kulkarni (1997) who discovered an unusually subluminous 
($L_x \approx  10^{-2} L_{\rm Edd}$) ``Type I'' X-ray burst from the direction 
of M28, thus leading one to expect the presence of one or more LMXBs.

In this paper we report on the first deep X-ray observations of M28 using  
the ACIS-S detector aboard the \CXO.
We show that the X-ray core emission seen with the ROSAT HRI is dominated
by a superposition of multiple discrete sources; we measure the unconfused 
phase-averaged spectrum of the X-ray flux from the 3.05-ms PSR B1821$-$24; we 
establish the X-ray luminosity function down to a limit of about $6 \times  
10^{31}$ erg s$^{-1}$ and measure, with high precision, the absolute positions 
of all the detected X-ray sources in M28 to facilitate identifications in 
other
wavelength bands. Observations and  data analysis are described in \S2.1 
through
\S2.3.
In  addition, we  have used archival Hubble Space Telescope ({\sl HST}) 
observations to search for potential optical counterparts of the sources
detected by the \CXO\ (\S3).

\section{Observations and data analysis \label{obs}}

M28 was observed three times for approximately equal observing intervals of 
about 13 ksec between July and September 2002 (Table~\ref{observations}). 
These observations were scheduled so as to be sensitive to time variability 
on time scales up to weeks.  The observations were made using 3 of the 
\CXO\ Advanced CCD Imaging Spectrometer (ACIS) CCDs (S2,3,4) in the faint 
timed exposure mode with a frame time of 3.241 s. Standard {\sl Chandra} 
X-Ray Center (CXC) processing (v.6.8.0) has applied aspect corrections and 
compensated for spacecraft dither. Level~2 event lists were used in our 
analyses. Events in pulse invariant channels corresponding to 
$\approx 0.2$ to 8.0 keV were selected for the purpose of finding sources. 
Due to uncertainties in the low energy response, data in the range 0.5 to 8.0
keV 
were used for spectral analyses.  Increased background corrupted a small 
portion 
of the third data set reducing the effective exposure time from 14.1 ksec to
11.4 
ksec (Table~\ref{observations}) although no results were impacted by the
increased 
background.

The optical center of the cluster at $\alpha_{2000} = 18^{\rm h}\,24^{\rm
m}\,32\fs89$  
and $\delta_{2000} = -24\degr\, 52\arcmin\, 11\farcs4$ (Shawl and White 1986)
was 
positioned 1\arcmin\ off-axis to the nominal aim point on the back-illuminated
CCD, 
ACIS-S3, in all 3 observations. A circular region with 3\farcm1-radius,
corresponding 
to twice the half-mass radius of M28, centered at the optical center was
extracted 
from each data set for analysis.  No correction for exposure was deemed
necessary 
because the small region of interest lies far from the edges of the S3 chip.

The X-ray position of \psr\ was measured separately using the three data sets
and the 
merged data. The results of these measurements are listed in
Table~\ref{psr_positions}. 
The set-averaged position is the same as that derived using the merged data 
set. 
The root-mean-square (rms) uncertainty in the pulsar position, based on the 3
pointings, 
is 0\farcs042 in right ascension and 0\farcs029 in declination.  The radio
position and 
proper motion of the pulsar, as measured by Rutledge et al.\ (2003), places 
the
pulsar  
at the time of the observation only $\Delta_{\alpha}=$ 0\farcs083,
$\Delta_{\delta}=-0\farcs042$ 
away from the best-estimated X-ray position.  In what follows the observed 
X-ray
positions 
of all sources have been adjusted to remove this offset.

\subsection{Image Analysis \label{image_analysis}}

\begin{figure}
 \centerline{\psfig{figure=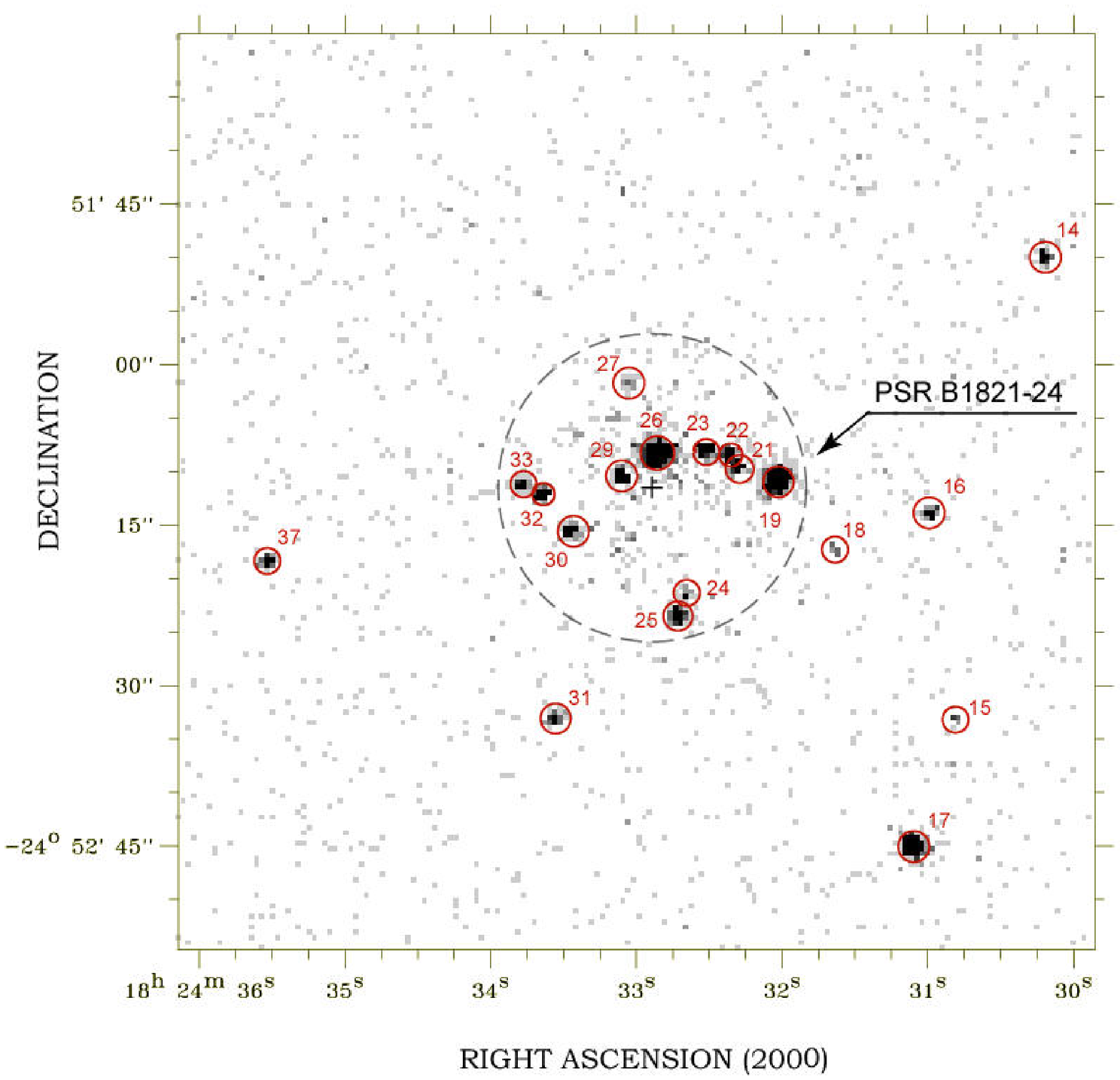,width=16cm,clip=}}
 \caption[]{
\CXO\ ACIS-S3 image of the central region of M28. Twelve X-ray sources
are detected within 
the 0\farcm24 core-radius, indicated by a dashed circle. The optical center of
the cluster is 
indicated by a cross. The X-ray counterpart of the millisecond pulsar 
B1821--24
is source \#19.}
\label{3Rc_image}
\end{figure}

The central portion of the combined \CXO\ image is shown in
Figure~\ref{3Rc_image}. 
We used the same source finding techniques as described in Swartz et 
al.~(2002)
with 
the circular-gaussian approximation to the point spread function, and a 
minimum 
signal-to-noise (S/N) ratio of 2.6 resulting in much less than 1 accidental
detection 
in the field. The  corresponding  background-subtracted  point source 
detection
limit is 
$\sim$10 counts. The source detection process was repeated using the CXC 
source
detection 
tool  {\tt wavdetect} (Freeman \etal\ 2002) and yielded consistent results at
the  equivalent 
significance level. Forty-four sources were found using these detection
algorithms.
Close inspection of the image  with the source positions overlaid and of the
source time 
variations showed that one source detected by the software was really two
sources 
(numbers 21 and 22  of Fig.\ \ref{3Rc_image}), and that an additional source, 
\#
24, is also present.

Table~\ref{M28_source_table} lists the 46 X-ray sources. The table gives the
source positions, 
the associated uncertainty in these positions, the radial distance of the 
source
from the 
optical center, the signal-to-noise ratio, the aperture-corrected counting 
rates
in various 
energy bands, the unabsorbed luminosity in the $0.5-8.0$ keV energy range 
based
on the spectroscopy 
discussed in \S~\ref{spectral_analysis}, and a variability designation 
according
to the discussion 
in \S~\ref{time_variability}.

The positional uncertainty listed in column~4 of Table~\ref{M28_source_table} 
is
given by 
$r=1.51(\sigma^2/N + \sigma_o)^{1/2}$ where $\sigma$ is the size of the 
circular
gaussian that 
approximately matches the PSF at the source location, $N$ is the
aperture-corrected number of 
source counts, and $\sigma_o$ represents the systematic error. Uncertainties 
in
the plate 
scale\footnote{see http://asc.harvard.edu/cal/Hrma/optaxis/platescale/} imply 
a
systematic 
uncertainty of $0\farcs13$, and, given that the radio and \CXO\ positions 
agree 
to $0\farcs2$, we feel that using $0\farcs2$ is a reasonable and conservative
estimate for 
$\sigma_o$. The factor 1.51 is the radius that encloses 68\% of the circular
gaussian. 
The parameter $\sigma$, varies from $\sim0\farcs9$ near the aimpoint to  
$<2\farcs1$ near the edge of the $3\farcm1$-radius extraction region.

The rates in the soft, medium, and hard bands listed in
Table~\ref{M28_source_table} are 
background subtracted and have (asymmetrical)  67\% confidence uncertainties
based on 
Poisson statistics (rather than the usual symmetrical Gaussian approximation).
When the band-limited rates are positive, the uncertainties are symmetrical in 
probability space (that is we set the lower and upper limits to the 67\%
confidence 
interval so that the true source rate is equally likely to fall on either side
of 
the derived rate), but not in rate space. Otherwise, we set the rate to zero 
and 
calculate a 67\% upper limit.

From Figure~\ref{3Rc_image} we see that there are 12 point sources in the
central region in the summed image. In addition, there remains some unresolved 
emission from this region of the cluster 
(\S~\ref{spectroscopy_central_region}).
Pooley et al. (2002b) found similar diffuse emission in the central regions of 
the \CXO\ image of the globular cluster NGC 6440.

\subsubsection{Radial Distribution}

The projected surface density, $S(r)$, of detected X-ray sources was compared 
to a King profile, $S(r)=S_o [1+(r/r_o)^2]^{-\beta} + C_0$. The constant term 
$C_0$ was added to account for background sources. We estimated the number of 
background sources using our observed flux limits of $\sim5.5 \times 
10^{-16}$~\ergcms\ in the 0.5--2.0 keV band and $\sim1.5 \times
10^{-15}$~\ergcms\ 
in the 2.0--10 keV band. These limits, together with the $\log N(>S)$-$\log S$ 
distribution from the \CXO\ Deep Field South (Rosati et al.\ 2002) gives 
an estimate of at least $\sim$0.3 to 0.4 background sources per square arcmin 
in 
the field. (The true value may be higher because of the low galactic latitude 
of
M28
relative to the deep field.) The best-fit value for $C_o$ is $0.36\pm0.22$
sources 
per arcmin$^2$ indicating that there are $\sim10$ sources in the
$3\arcmin$-radius 
field not associated with the cluster. 
The other fit parameters ($S_o = 123\pm11$ sources per arcmin$^2$, $r_0  =
23\farcs8^{+10.8}_{-5.9}$,\ $\beta = 3.51^{+2.7}_{-1.1}$) can be used to
estimate the core radius, $r_{c,x}$, and the typical mass, $M_x$, of the X-ray
source population. 
We find a best fit core radius $r_{c,x} = 10\farcs9^{+8.8}_{-4.7}$.
This is comparable to the distribution of optical light of the cluster:
$r_{c\ast}=14\farcs4$ (Harris 1996). Following the derivation of Grindlay et 
al.\ (2002), the best-fit mass of the X-ray sources is $M_x =
1.87^{+1.25}_{-0.49}$~\msun, assuming the dominant  visible stellar population
has a mass of $M_{\ast} \sim 0.7$~\msun.
Although our range for $M_x$, estimated in this way, barely overlaps the  
range $1.1-1.4$~\msun\ deduced by Grindlay et al. \ (2002) for 47~Tuc,
additional effects due to uncertainties in cluster properties (position of  
cluster center, core radius, mass segregation, etc.) mean that the two  
results are in fact indistinguishable.

\subsection{Spectral Analysis\label{spectral_analysis}}

Point-source counts and spectra were extracted from within radii listed in 
Table~\ref{bright_5_t1}. Because the field is so crowded, background was
estimated
using a region of $\sim 50\arcsec$ radius located in the south-western portion 
of the field. The background rate, used for all calculations, was
$3.7\times10^{-6}$
counts s$^{-1}$ arcsec$^{-2}$.

Only 6 of the 46 detected sources have sufficient counts to warrant an
individual spectral analysis. In descending order of the number of detected 
counts these are sources \#26, \#19, \#4, \#17, \#25, and \#28.
Source \#19 is the X-ray counterpart of the PSR B1821--24.
The results of fitting various spectral models to the energy spectra of the 
brightest sources are presented in \S~\ref{psr}, \S~\ref{brt26}, and
\S~\ref{brt}. All spectral analyses used the CXC CALDB 2.8 calibration 
files (gain maps, quantum efficiency uniformity and effective area).
The abundances of, and the cross sections in, TBABS (available in XSPEC 
v.11.2) 
by Wilms, Allen \& McCray (2000) were used in calculating the impact of the 
interstellar absorption. All errors are extremes on the single interesting 
parameter 90\% confidence contours.

A correction\footnote{available from
http://heasarc.gsfc.nasa.gov/docs/software/lheasoft/xanadu/xspec/models/acisabs.html}  
was applied when fitting models to the spectral data to account for the 
temporal
decrease in low-energy sensitivity of the ACIS detectors due, presumably, to 
contamination buildup on the ACIS filters.
The correction was based on the average time of the observations after
launch of 1105 days.

For the remaining 40 sources, all with fewer than 100 detected source counts, 
the spectra were combined to determine the mean spectral shape and total
luminosity. 
Fits were made using both an absorbed power law ${\rm d}N/{\rm d}E \propto  
E^{-\Gamma}$ (best-fit $\Gamma = 1.73\pm 0.18$, $N_{22} = N_H/(10^{22}\, {\rm
cm}^{-2}) = 0.18 \pm 0.06$, $\chisq = 89.8$ for 86 degrees of freedom (dof))
and an absorbed thermal bremsstrahlung model ($kT=7.40\pm 2.10$~keV, $N_{22}=
0.12\pm0.04$, $\chisq=92.8$). 
Using the power-law model parameters, the total (unabsorbed) flux from the 40
weak sources is $2.34\pm0.11\times 10^{-13}$ \ergcms\ in the 
0.5--8.0 keV band, and the corresponding X-ray luminosity is $8.47 \pm 0.40 
\times 10^{32}$ \ergl. 
This spectrum was then used to estimate the individual source luminosities of
the 40 faint sources listed in Table~\ref{M28_source_table}. 
See also the discussion in \S~\ref{color-l}.

\subsubsection{The Phase-averaged Spectrum of PSR 1821--24 \label{psr}}

\begin{figure}
 \centerline{\psfig{figure=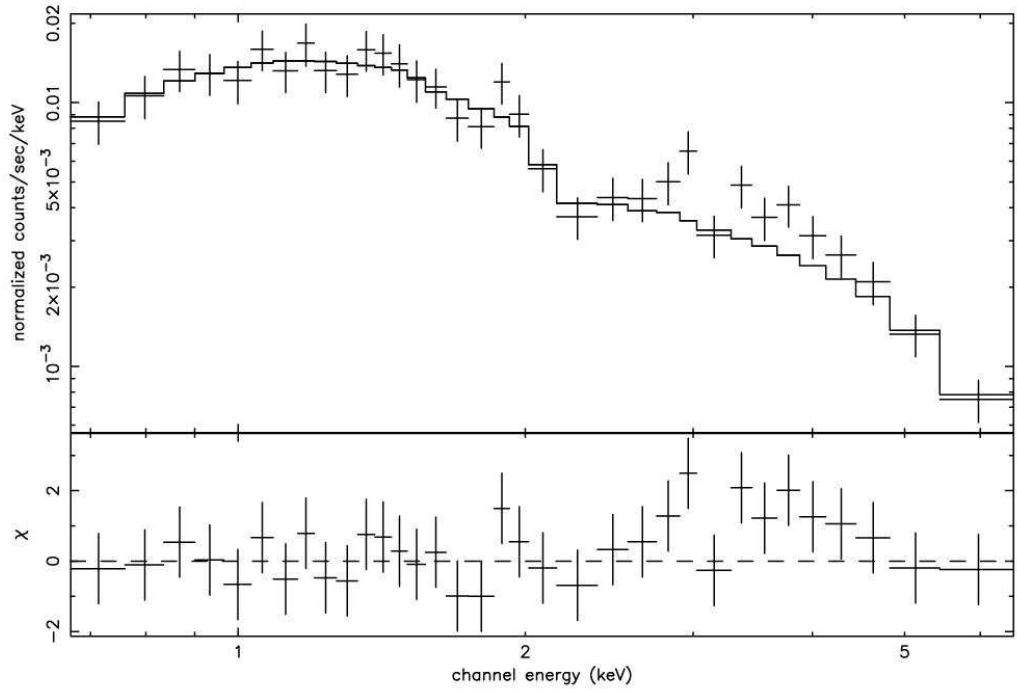,width=14cm,clip=}}
 \caption[]{Energy spectrum of the millisecond pulsar \psr\ fit to an
absorbed power-law model ({\it upper panel}) and contribution to the  
$\chisq$ fit statistic ({\it lower panel}).
The residuals indicate a marginally significant feature near 3 keV.
}
\label{psr_spectrum}
\end{figure}

The spectrum of \psr\  was measured  by extracting  $\sim1100$ counts within a
radius  of 1\farcs72 centered on the pulsar position. 
A background subtraction was performed, but its contribution ($\sim 2$ counts)
is negligible.
Neither our source-detection algorithm nor {\tt wavdetect} found any evidence
for a spatial extent to the pulsar's X-ray counterpart; thus, using the
CXC model point spread function, 97\% of all the events from
\psr\  are within the selected region.
The data were binned into 34 bins guaranteeing at least 30 counts per bin.
Model spectra were then compared with the observed spectrum.
A power-law model was found to give a statistically adequate
representation of the observed energy spectrum; the best-fit spectrum and
residuals are shown in Figure~\ref{psr_spectrum}.
A thermal bremsstrahlung model resulted in a slightly better fit, but is
not considered to be physically applicable.
A blackbody model does not fit the data ($\chisq =110$ for 31 dof),
as one could expect based upon the hardness of the observed spectrum
and the similarity of the sharp X-ray and radio pulse profiles  (see, for
example, Becker \& Pavlov 2001, Becker \& Aschenbach 2002).

The best-fit power law yields $N_{22}=0.16^{+0.07}_{-0.08}$,
$\Gamma = 1.20^{+0.15}_{-0.13}$, and a normalization of
$3.74 ^{+1.0}_{-0.48}\times10^{-5}$ photons cm$^{-2}$ s$^{-1}$ keV$^{-1}$ at
$E=1$ keV ($\chi_\nu^2= 0.89$ for 31 dof).
The column density is in fair agreement with what is deduced from the 
reddening
towards M28. The unabsorbed energy flux in the 0.5--8.0 keV band is
$f_x= 3.54^{+0.06}_{-0.05}\times 10^{-13}\,\,{\rm ergs\,\, s}^{-1}\,{\rm
cm}^{-2}$, yielding an X-ray luminosity of $L_x= 1.28\pm0.02\times 10^{33}\,
{\rm ergs\,\, s}^{-1}$. 
This luminosity implies a rotational energy to X-ray energy conversion factor
$L_x/\dot{E}= 5.8\times 10^{-4}$. If transformed to the ROSAT band, this 
corresponds to $L_x= (3.4$--$4.0) \times 10^{32}\,{\rm ergs\,\, s}^{-1}$, and 
is 
similar to the luminosity inferred from the ROSAT data (Verbunt 2001).
The photon index we found is compatible with $\Gamma\sim 1.1$ deduced
for \psr\ from the observations at pulse maximum using {\sl RXTE} data
(Kawai \& Saito 1999). We note that we have ignored the possible effects of 
photon pileup, and this could artificially harden the spectral index.
However, the degree of pileup here is sufficiently small ($< 0.12$ counts
per frame), so that its effect on the spectrum is not significant.
In fact, application of the Davis (2001) pileup model suggests
the spectral index would be steeper by only ~0.1 in the absence of pileup.

The residuals in Figure~\ref{psr_spectrum} hint at a spectral feature or
features at an energy slightly above 3 keV. Although the reduced 
number of counts made it impossible to determine
whether  the feature was  present in the three separate observations, we
note that some  excess emission in the band between 3 and 4 keV is seen in
all three data sets. By adding a gaussian ``line'' to the power law model,
we found a line center at 3.3 keV with a gaussian width of 0.8 keV and a 
strength of $\approx 6\times 10^{-6}$ photons cm$^{-2}$ s$
^{-1}$, which corresponds to a luminosity of $\approx 1.1\times 10^{31}\,\, 
{\rm
erg\, s}^{-1}$.
Adding the line changes $\chi_\nu^2$ to 0.63 (for 28 dof).
The F-test indicates that the addition of this line component is
statistically significant at 98\% confidence, i.e., evidence for a broad
spectral feature is marginal.

If we assume that this feature is real, then it is interesting to speculate as
to its origin.
There are some atomic lines, from K and Ar, close to the 3.3 keV energy.
However, K is not an abundant element, and it is hard to explain why
the Ar lines are observed while no lines are seen from other, more
abundant, elements. Therefore, we consider the more likely possibility 
that this is an electron cyclotron line, formed in a magnetic field $B\approx
3\times 10^{11}$ G.
Such a strong field can be explained by the presence of
either multipolar components or a strong off-centering of the magnetic
dipole or both.

For instance, if a dipole with the magnetic moment
$\mu = 2\times 10^{27}$ G cm$^3$ (see \S1) is shifted along its axis
such that it is 1.9 km beneath the surface,
the magnetic field at the closest pole is $3\times 10^{11}$ G.
The cyclotron line could be formed in an optically thin, hot corona above
the pulsar's polar cap, with a temperature $kT\sim 10$ keV such that the
line's Doppler width is smaller than observed while the excited Landau
levels are populated.
The observed luminosity in the line can be provided by as few as
$N_e = 2.4\times 10^{25}\, T_8^{-1} B_{11.5}^{-2}$ electrons,
where $T_8=T/(10^8\,{\rm K})$, $B_{11.5}=B/(3\times 10^{11}\,{\rm G})$.

The corona is optically thin at the line center for an electron column density
$n_e H < 5\times 10^{18}\, T_8^{1/2} B_{11.5}$ cm$^{-2}$ (where $H$ is the
geometrical thickness of the corona) and a polar cap area $A_{\rm
pc}>4.6\times 10^6\, T_8^{-3/2} B_{11.5}^{-3}$ cm$^2$ (polar cap radius
$R_{\rm pc} > 12$ m).
The standard estimate of the polar cap radius, applied to the off-centered
dipole, gives $R_{\rm pc} \sim r (2\pi r/cP)^{1/2} = 220$ m, where $r$ ($=1.9$
km in our example) 
is the radial distance from the center of the magnetic dipole to the surface.
At this value of $R_{\rm pc}$, the optical thickness at the line center is
$\tau \sim 0.003\, T_8^{-3/2} B_{11.5}^{-3}$.
If the corona is comprised mainly of a proton-electron plasma, then its
thickness can be estimated as $H \sim kT\, (m_H g)^{-1}\sim 40\,T_8$ cm
(at $g\sim 2\times 10^{14}$ cm s$^{-2}$, typical for a neutron star), and
a characteristic electron number density is $n_e\sim 5\times 10^{14}\,
T_8^{-2} B_{11.5}^{-2}$ cm$^{-3}$.
Thus, the cyclotron interpretation of the putative line looks quite plausible,
and confirming the line with deeper observations would provide strong evidence
of local magnetic fields at the neutron star surface well above the
``conventional'' magnitudes inferred from the assumption of the centered 
dipole
geometry.

\subsubsection{Source \#26: An LMXB in quiescence?\label{brt26}}

Of particular interest among the brightest  X-ray sources detected in M28  
is the luminous, soft source \#26. We extracted data as per 
Table~\ref{bright_5_t1} and fit it with various spectral models.
As expected from the softness of the count spectrum (Fig.~\ref{bright_26}),
the power-law fit yields a photon index, $\Gamma\approx 5.2$,
well in  excess of those observed from known astrophysical sources with
power-law spectra, and the hydrogen column density, $N_{22}\approx 0.68$,
that significantly exceeds the values expected from the M28's reddening (\S1)
and measured for the pulsar (\S2.2.1). Therefore, we tried various models 
of thermal radiation. A blackbody fit gives the hydrogen column density 
$N_{22}\approx 0.13$,  temperature $kT_{\rm BB}\approx 0.26$ keV
and radius  $R_{\rm BB}\approx 1.3$ km, corresponding to the bolometric
luminosity of $L_{\rm BB}\sim 1\times 10^{33}$ erg s$^{-1}$.
Such values are typical for blackbody fits of LMXBs with transiently
accreting neutron stars in quiescence (e.g., Rutledge et al.\ 2000).
The relatively high temperatures of such old neutron stars can be
explained by heating of the neutron star crust during the repeated
accretion outbursts (Brown, Bildsten, \& Rutledge 1998).
This heat provides an emergent thermal luminosity $L=8.7\times 10^{33}
\langle \dot{M}\rangle_{-10}$ erg s$^{-1}$, for the nuclear energy release of
1.45 MeV per accreted nucleon, where $\langle \dot{M}\rangle_{-10}$ is the 
time-averaged accretion rate in units of $10^{-10}\, M_\odot\,\,{\rm 
yr}^{-1}$.

The radius $R_{\rm BB}$ obtained from the blackbody fit is much smaller than 
$R\sim 10-15$ km expected for a neutron star. A likely reason for this
discrepancy 
is that the blackbody model does not provide an adequate description of 
thermal 
radiation from the neutron star surface. At the temperatures of interest the 
neutron star crust is covered by an atmosphere comprised of the accreted 
matter.
Because of the gravitational sedimentation, the outermost layers of such an 
atmosphere, which determine the properties of the emergent radiation, are  
comprised of hydrogen, the lightest element present.
Since the magnetic fields of neutron stars in LMXBs are expected to be
relatively low, $\la 10^{9}$ G, they should not affect the properties of
X-ray emission, which allows one to use the nonmagnetic hydrogen atmosphere  
models (Rajagopal \& Romani 1996; Zavlin, Pavlov \& Shibanov 1996).
We fit the observed spectrum with the {\tt nsa} model\footnote{These models
are based on the work of Zavlin et al.\ (1996), with additional physics to  
account for comptonization effects (Pavlov, Shibanov \& Zavlin 1991)}
in XSPEC (v.11.2). In applying this model, we set the neutron star mass 
to 1.4\msun, leaving  the radius of the emitting region and the surface 
temperature as free parameters. We obtained a statistically acceptable 
fit ($\chi_\nu^2=0.96$, $\nu=44$), with the hydrogen column density 
$N_{22}=0.26\pm 0.04$ (consistent with the expected value), the effective 
temperature $kT_{\rm eff}^\infty = 0.09^{+0.03}_{-0.01}$ keV,
a factor of 3 lower than $T_{\rm BB}$, and the radius $R^\infty =
14.5^{+6.9}_{-3.8}$ km, comparable with a typical neutron star radius.
(The superscript $^\infty$ means that the quantities are given as
measured by a distant observer; they are related to the quantities as measured
at the neutron star surface as $T^\infty = g_r T$, $R^\infty = g_r^{-1} R$,
$L_{\rm bol}^\infty = g_r^2 L_{\rm bol}$, where $g_r = (1-2GM/Rc^2)^{1/2}$ 
is the gravitational redshift factor. The values of $T_{\rm eff}^\infty$ 
and $R^\infty$, as inferred from the atmosphere model fits, are considerably 
less sensitive to the assumed value of neutron star mass.)

The corresponding bolometric luminosity is $L_{\rm bol}^\infty =
1.9^{+1.1}_{-0.6}\times 10^{33}$ erg s$^{-1}$. Such a luminosity can be 
provided by a time-averaged accretion rate of $\langle\dot{M}\rangle \sim
2\times 10^{-11}\,M_\odot\,\,{\rm yr}^{-1}$. Because all the fitting 
parameters 
we obtained for source \#26 are typical for quiescent radiation of
other transiently accreting neutron stars in LMXBs (e.g., Rutledge et al.\  
2002b, and references therein), we conclude that such an interpretation is  
quite plausible.

In addition to the thermal (photospheric) component, some X-ray transients  
in quiescence show a power-law high-energy tail (Rutledge et al.\ 2002b),  
apparently associated with a residual low-rate accretion. We can see from 
Figure~\ref{bright_26} that our thermal model somewhat underestimates the 
measured flux above 2.5 keV. This is a consequence of a mild (0.16 counts 
per frame) amount of pileup. The pileup model suggested by Davis (2001), 
as implemented in  XSPEC v.11.2, does reproduce this tail, but only marginally 
changed the best-fit parameters.

The quiescent emission of some of transient LMXBs shows appreciable variations
of X-ray flux, with a time scale of a month (Rutledge et al.\ 2002b), and our
source \#26 also appears to be time variable (\S\ref{time_variability}) in the
three observations. We performed spectral fits to each of the data sets
separately 
to check for spectral variation and found none, indicating that the time
variability (\S~\ref{time_variability}) is not dominated by spectral
variations.

To test alternative interpretations of source \#26, we also fit various
model spectra of an optically thin thermal plasma in collisional equilibrium,
applicable to stellar coronae and similar sources. Fitting the spectrum with 
the XSPEC model {\tt mekal}\footnote{This and similar models include thermal
bremsstrahlung and line emission as components. Such models become equivalent 
to the optically thin thermal bremsstrahlung at high temperatures and/or low 
metalicities, while they strongly differ from the bremsstrahlung in the 
opposite 
case, when the main contribution to the soft X-ray range comes from the line 
emission. Therefore, there is no need to consider the thermal bremsstrahlung 
fits if more advanced models for optically thin plasma are available for
fitting.},
for $Z=0.02 Z_\odot$, we obtain a good fit ($\chi_\nu^2=0.88$
for 44 dof) with $N_{22}\approx 0.33$, $kT \approx 0.6$ keV,
and emission measure EM $\approx 2\times 10^{57}$ cm$^{-3}$, at $d=5.5$ kpc.
The corresponding luminosity, $L_x(0.5- 8\, {\rm keV}) \approx 1.2\times  
10^{33}$ erg s$^{-1}$, strongly exceeds the maximum luminosities of coronal 
emission observed from either single or multiple nondegenerate stars of any 
type available in an old globular cluster. The spectrum is too hard, and the 
luminosity is too high, to interpret this emission as produced by a 
non-magnetic 
CV (Warner 1995, and references therein). The inferred temperature is too low 
in comparison with the typical temperatures, $\sim 30-40$ keV, of the hard 
X-ray (bremsstrahlung) component observed in  polars (magnetic CVs, in which 
rotation of the accreting magnetic white dwarf is synchronized
with the orbital revolution), and, in addition, this component is usually 
much less luminous in polars. On the other hand, the spectrum of source 
\#26 is too hard to be interpreted as a soft X-ray component
($20-40$ eV blackbody plus cyclotron radiation) observed in many polars.
Luminosities up to $\sim 1\times 10^{33}$ erg s$^{-1}$ (in the $2-10$ keV 
band)
have been observed in a number of intermediate polars (asynchronous magnetic 
CVs 
with disk accretion). However, spectra of intermediate polars are, as a rule,
much harder (similar to those observed in polars) and strongly absorbed
($N_{22} \sim 10$) by the accreting matter (Warner 1995).
Therefore, we conclude that the interpretation of source \#26 as a stellar
corona or a CV looks hardly plausible, and most likely its X-ray emission  
emerges from the photosphere of a quiescent transient neutron star in an LMXB.

\subsubsection{Spectra of Four Other Bright Sources}\label{brt}

In addition to \psr\ and source \#26, there are 4 moderately bright sources 
with enough counts to attempt spectral fitting. Data from an extraction radius 
large enough to encompass a significant fraction of the source counts were
binned 
into spectral bins to maintain a minimum number of counts per bin, background 
subtracted, and fit to various spectral models. The extraction radii, total
number 
of extracted counts, estimated number of background counts, minimum number of 
counts per spectral bin, and the number of spectral bins are listed in 
Table~\ref{bright_5_t1}. To characterize the spectra of these sources, we fit 
each of them with three popular models of substantially different shapes: 
a black-body, a power-law, and an optically-thin thermal emission model 
({\tt mekal}) with $Z=0.02\,Z_\odot$. The best-fit parameters for these 
models, 
including uncertainties, are given in Table~\ref{bright_5_t2}.

The brightest of the four sources is source \#4. Its spectrum (Fig.\ 4) is 
too hard to consider it as a qLMXB. The power-law fit of the spectrum, with 
$\Gamma\approx 1.6$, might be interpreted as arising from magnetospheric 
emission from an MSP, with a luminosity $L_x\approx 9\times 10^{32}$ erg
s$^{-1}$, 
in the $0.5-8$ keV range. However, the hydrogen column density inferred from 
the power-law fit, $N_{22}\approx 0.86$, considerably exceeds those estimated 
from the interstellar reddening and the power-law fit of the \psr\ spectrum.
The large $N_H$ and a large distance, 2\farcm53 from the cluster's center,
together with the power-law slope typical for AGNs, hint that source \#4
could be a background AGN (notice that the Galactic HI column density in
this direction is about $0.19\times 10^{22}$ cm$^{-2}$ --- Dickey \&
Lockman 1990). The blackbody fit gives $N_{22}\approx 0.16$, consistent 
with that obtained for \psr. This fit indicates that source \#4 might be a 
thermally emitting MSP, although the blackbody temperature, $\approx 1$ keV, 
is surprisingly high, and the blackbody radius, $\approx 60$ m, is much 
smaller than expected for a pulsar polar cap (a standard polar cap radius 
is $R_{\rm pc}= 1.4\,R_6^{3/2} P_{-2}^{-1/2}$ km, assuming a centered dipole, 
where $R_6$ is the neutron star radius in units of 10 km, and $P_{-2}$ is 
the pulsar's  period in units of 10 ms).
The inferred temperature would become lower by a factor of two, and the
radius would increase by an order of magnitude if we assume that this
emission emerges from a polar cap covered by a hydrogen or helium
atmosphere (e.g., Zavlin \& Pavlov 1998), but still a temperature of a few  
million kelvins looks too high, and a radius of a few hundred meters is 
somewhat too small, for a typical MSP. If the possible variability of 
this source (see Table 3) is confirmed by future observations, the 
interpretation of this source as an MSP can be ruled out.
The {\tt mekal} fit gives a temperature, $kT\sim 30$ keV, and a  
luminosity, $L_x(0.5-8\,{\rm keV})\sim 8 \times 10^{32}$ erg s$^{-1}$,  
too high to be interpreted as coronal emission of single or binary (BY   
Dra, RS CVn) nondegenerate stars.
The high temperature\footnote{At such high temperatures the {\tt mekal}  
model is essentially equivalent to the optically thin thermal  
bremsstrahlung.} is typical of CVs, and the high absorption column,  
$N_{22}\approx 0.77$, could be interpreted as additional absorption by the  
accreting matter, but the luminosity is somewhat higher than observed for  
most CVs (cf.\ Grindlay et al.\ 2001). Thus, the spectral fits suggest 
that source \#4 is likely a background AGN, but they do not rule out the 
interpretation that it is a CV or an MSP.

\begin{figure}
\centerline{\psfig{figure=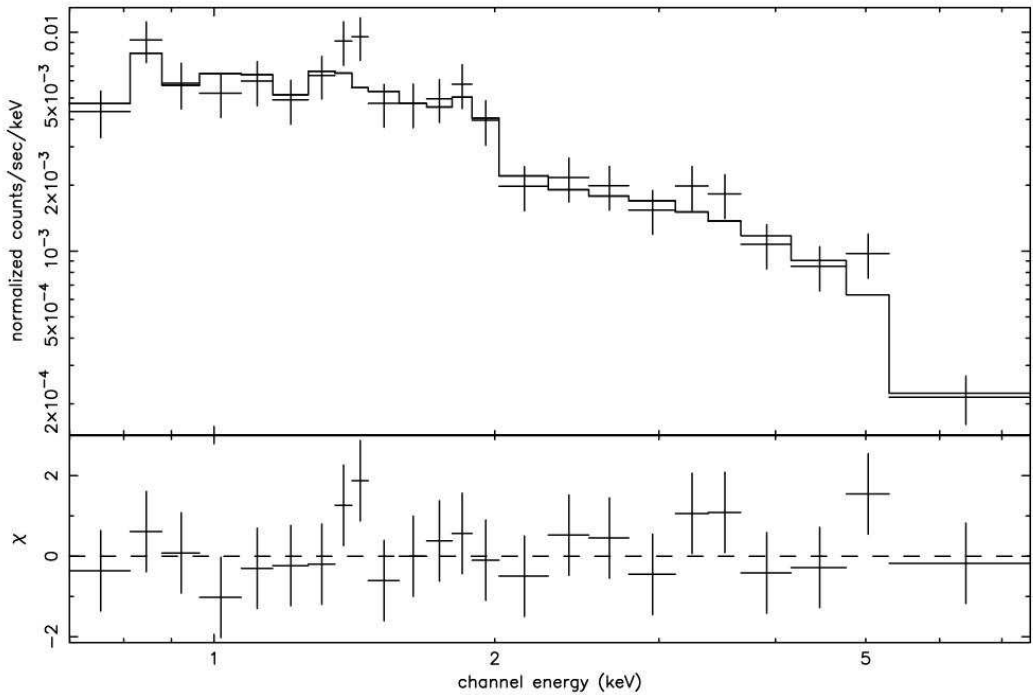,width=14cm,clip=}}
 \caption[]{Energy spectrum of source \#17, fit with an absorbed {\tt mekal}
model.}
 \label{bright_source_17_spectrum}
\end{figure}

The spectrum of source \#17 (Fig.~\ref{bright_source_17_spectrum}) is even  
harder than that of source \#4.
It can be equally well fitted with a power-law model, with
$\Gamma\approx 1.3$, and a {\tt mekal} model, with $kT\sim 40$ keV and  
$L_x(0.5- 8\,{\rm keV})\sim 5\times 10^{32}$ erg s$^{-1}$.
The blackbody fit does not look acceptable, because of the high $\chi_\nu^2
=1.73$ (the model underestimates the number of counts below 1 keV and  
above 5 keV) and unrealistically low $N_{22} < 0.03$.
The hardness of the spectrum is inconsistent with source \#17 being a qLMXB,
while its variability with a timescale of years (see \S2.3) rules out a  
MSP interpretation.
The luminosity of source \#17 is lower than that of \#4 so the  
argument against the CV interpretation is not so strong.
On the other hand, source \#17 is substantially closer to the cluster's  
center, so the probability that it belongs to the cluster is higher.  
Therefore, a CV interpretation looks more plausible.

\begin{figure}
\centerline{\psfig{figure=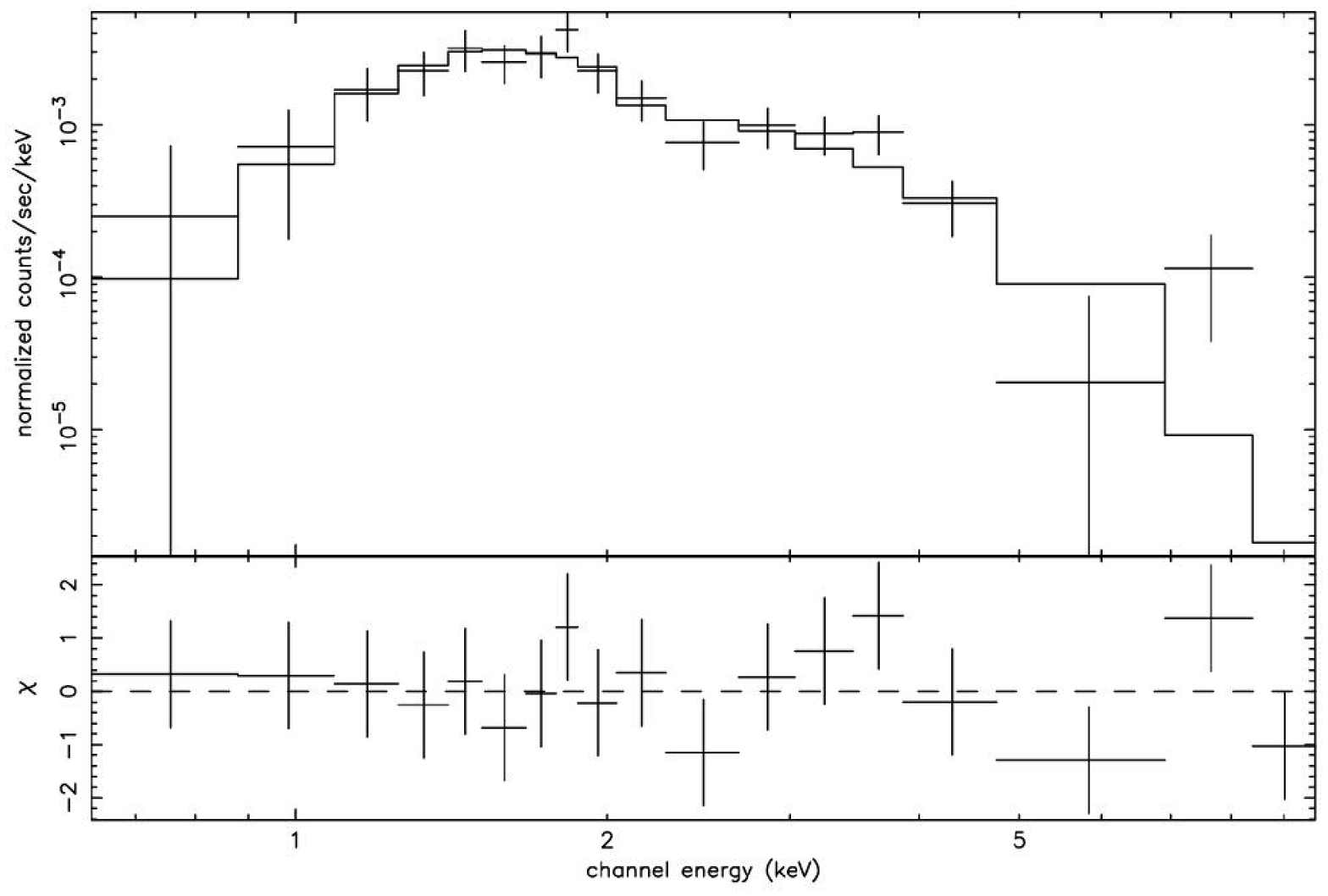,width=14cm,clip=}}
 \caption[]{Energy spectrum of source \#28, fit with an absorbed power-law
model.}
 \label{bright_source_28_spectrum}
\end{figure}

Source \#28 shows a softer spectrum (Fig.~\ref{bright_source_28_spectrum}),  
in comparison with sources  \#4 and \#17, with possible absorption or an  
intrinsic turnover at softer energies. Because of the small number of counts 
detected, we cannot distinguish between different fits statistically.
Both the power-law fit ($\Gamma\approx 3$, $N_{22}\sim 1.8$) and {\tt mekal}
($kT\approx 2$ keV, $N_{22}\sim 1.4$) require an absorption column much higher
than expected for a Galactic source in this direction. On the other hand, 
the blackbody fit yields a lower (albeit rather uncertain) absorption, 
$N_{22}\sim 0.3-1.1$. The blackbody temperature, $kT_{\rm BB}\sim 0.7$ keV, 
and the radius, $R_{\rm BB} \sim 100$ m, indicate that it might be thermal 
emission from an MSP  polar cap. As we have discussed for source \#4, a 
light-element atmosphere model would give a lower temperature, $kT\sim 0.3$ 
keV, 
and a larger radius,  $R\sim 1$ km, which makes the MSP interpretation even 
more plausible. The observed spectrum of source \#28 is harder than  
observed from qLMXBs,  so we consider the qLMXB interpretation unlikely.

\begin{figure}
\centerline{\psfig{figure=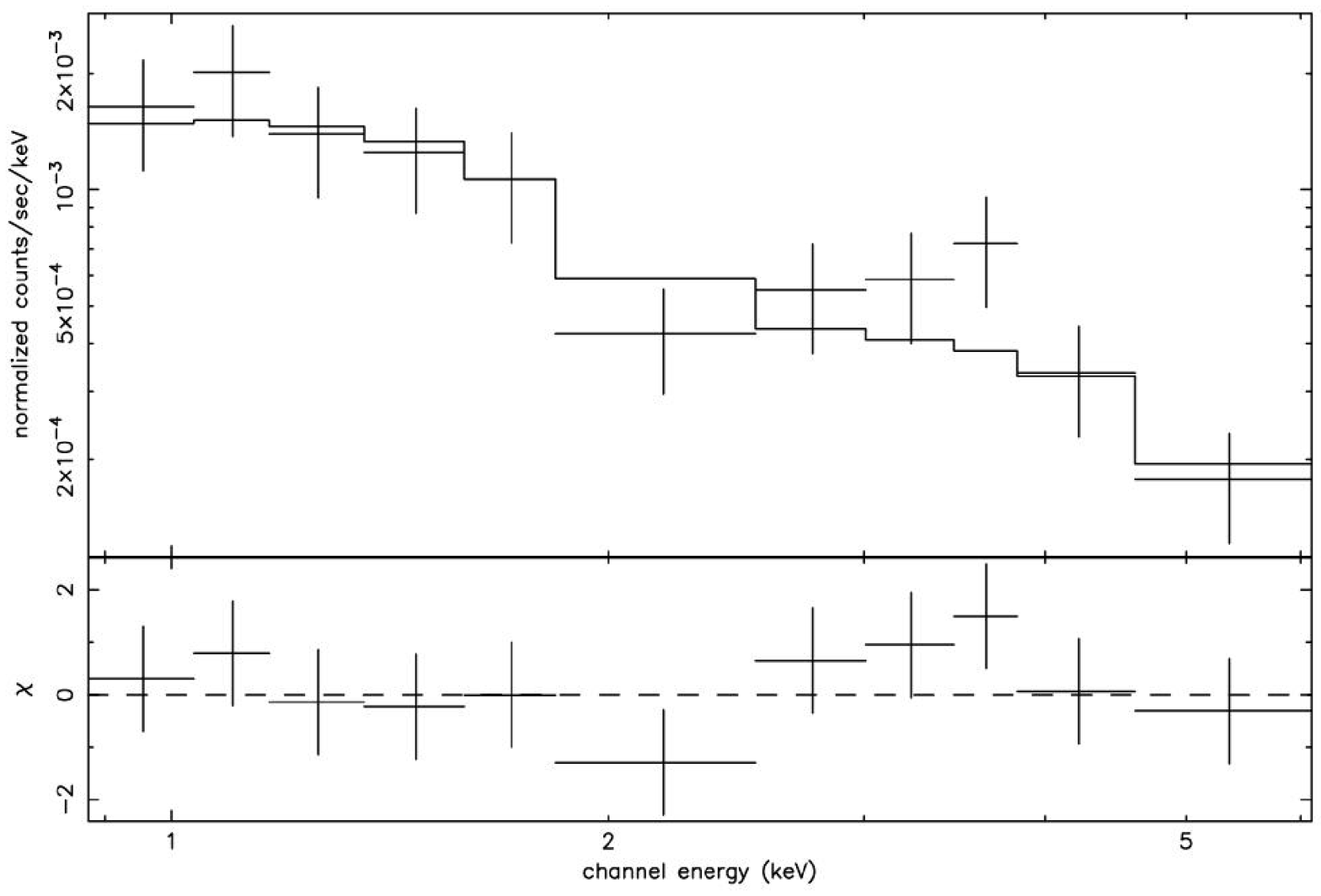,width=14cm,clip=}}
 \caption[]{Energy spectrum of source \#25, fit with an absorbed power-law
model.}
 \label{bright_source_25_spectrum}
\end{figure}

Finally, the hard spectrum of source \#25 
(Fig.~\ref{bright_source_25_spectrum}) 
strongly resembles that of source  \#17, although with much fewer counts.
The blackbody model is unacceptable, while both the power-law and {\tt  
mekal} yield reasonable fits. Similar to source \#17, we consider source 
\#25 as a plausible CV candidate.

\subsubsection{X-ray color-luminosity relation} \label{color-l}

\begin{figure}
 \centerline{\psfig{figure=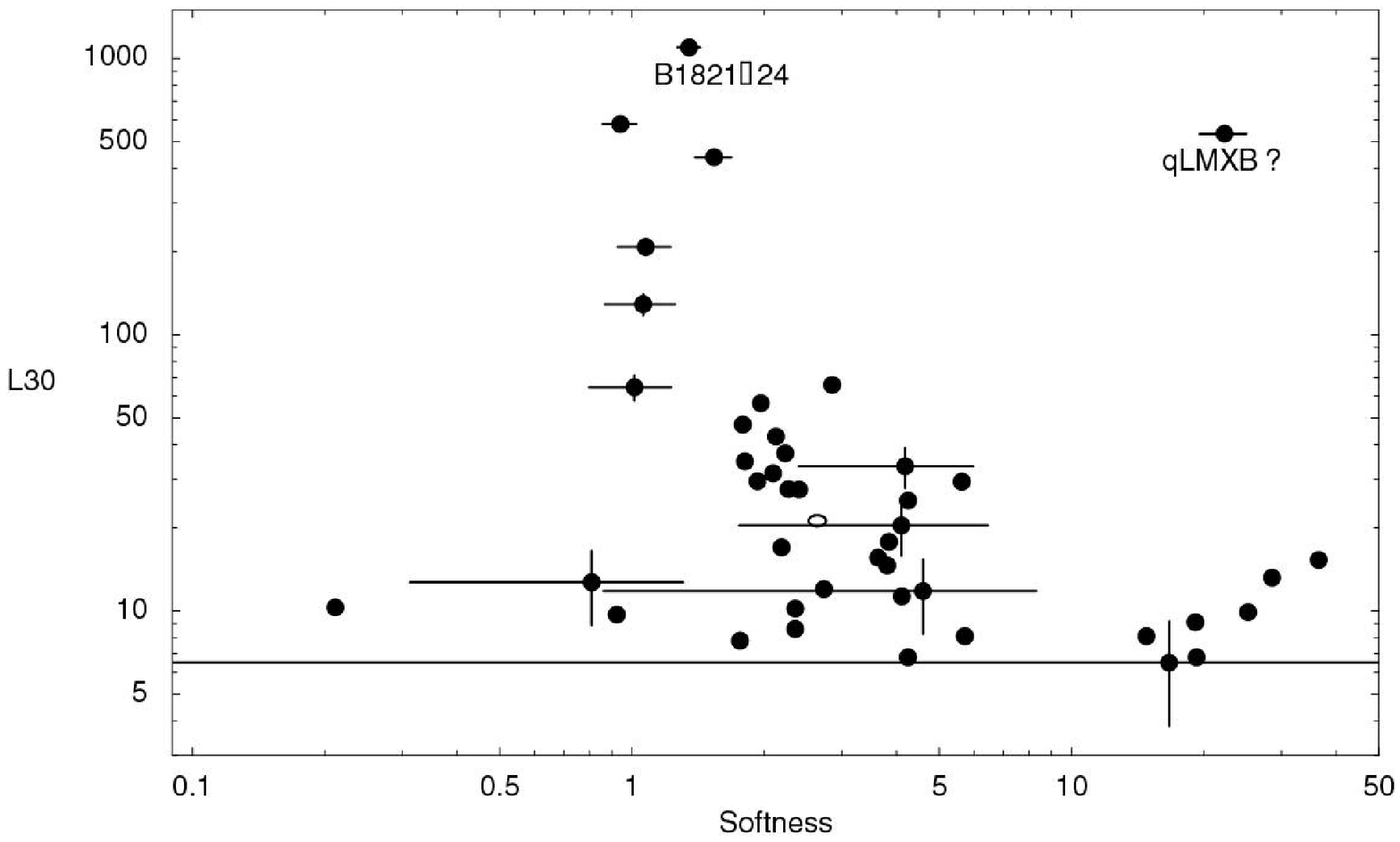,width=13cm,clip=}}
 \caption[]{X-ray luminosity (0.5--8.0 keV), in units of $10^{30}$ erg 
s$^{-1}$,
  vs.\ X-ray color for the sources listed in Table~\ref{M28_source_table}.
  The X-ray color (softness) is here defined as the rate in the 0.2--2.0 keV  
  band divided by the rate in the 2.0--8.0 keV band. The open circle marks the
softness for the average spectrum of the 40 faintest sources at their average
luminosity of $2.25 \times 10^{31}$ erg s$^{-1}$. Error bars 
  are displayed for selected sources. For the dimmest sources, X-ray color 
  is, practically speaking, unknown. The PSR B1821$-$24 is the most luminous 
  source in our sample. The source at the top right of the diagram, \#26, is 
  possibly a qLMXB.}
\label{cc_diagram}
\end{figure}

Figure \ref{cc_diagram} shows a plot of X-ray luminosity versus an X-ray
``color''. Such X-ray ``color-magnitude diagrams'' (CMDs; see Grindlay et 
al.~2001) are particularly useful for studying source populations in 
clusters where a large dynamic range of source luminosities and types 
can be studied at a common distance. The source with the highest luminosity, 
and hardest spectrum, is the PSR B1821$-$24. The source with the second 
highest luminosity we consider to be a good candidate for a quiescent 
low-mass X-ray binary (qLMXB) based on its spectral properties discussed 
in \S~\ref{brt}. The other sources presumably are a mix of CVs (especially 
those
with
luminosities $\gtrsim 10^{32}$ erg/s), RS CVns, main-sequence binaries,
MSPs, and other (unknown) systems.

Allowing for error bars on derived source colors and luminosities, it is
clear that the CMD is most useful for the classification of brighter
sources. However, even for faint sources, with poor statistics on each, a 
sufficiently large number of objects in the CMD can define an approximate 
distribution of source types (e.g. the MSPs in 47~Tuc) and constrain the 
possible source types of unidentified sources. In this observation, this 
has not been the case once the uncertainties are accounted for.
Nevertheless, it is interesting to speculate that many of the soft, faint  
sources are lower luminosity MSPs as in 47~Tuc.

\subsubsection{Spectroscopy of the central unresolved
emission}\label{spectroscopy_central_region}

As discussed in \S~\ref{image_analysis}, unresolved  X-ray emission is present
in the \CXO\ data.
This emission extends to roughly one core radius from the center of M28.
If we take the total counts detected within
$15''$ and subtract off the known contribution from point sources (including 
the
estimated counts from  
the full PSF) then we are left with an excess of about 541 counts.
To extract this spectrum, counts near the point sources were removed.
The resulting spectrum contained 568 counts in good agreement with our
estimate for the excess.
The background contribution to this total is $\sim$100 counts.
The spectrum was modeled using both a power-law model and a compound model  
consisting of a power-law with an additional optically-thin thermal
emission-line ({\tt mekal}) model.
To account for the underabundance with respect to solar expected in the
globular cluster, the abundance of metals in the {\tt mekal} model were set to 
2\% of their solar values.
The best-fitting model is the compound model ($\chisq = 30.4$ for 32
dof) though the {\tt mekal} component is significant at only the $\sim
2\sigma$ level ($\chisq = 37.3$ for 34 dof for the power-law-only model).
The best-fit photon index is $\Gamma = 1.79^{+0.37}_{-0.32}$, and the
temperature of the emission-line component is
$kT=0.18^{+0.11}_{-0.07}$~keV.
The luminosity
is $2.9^{+0.3}_{-0.3} \times 10^{32}$~\ergl\ ($6.0^{+1.1}_{-1.8} \times  
10^{32}$~\ergl\ after correction for absorption).

Interestingly, the photon index is similar (but not identical) to that
deduced from co-adding the 40 weakest resolved sources
(\S\ref{spectral_analysis}).
This suggests a portion of the unresolved emission may be from point sources
below the detection threshold but with similar spectral properties as
those above the detection threshold.

We find that the X-ray $\log N(>S)$-$\log S$ distribution of the 12
sources within 15$\arcsec$ of the cluster center is $N(>S) \sim 52\, 
S^{-0.53}$. 
Assuming that this relationship extends to lower counting rates, we can
estimate the number of photons that could come from sources below our
threshold of ~10 counts.
This extrapolation predicts that $\sim 200$ counts or $\sim 1.2\times
10^{32}$~\ergl\ is contributed by sources below the detection threshold.
Thus, unresolved sources with spectral properties similar to the weaker
resolved sources can account for roughly half the observed power-law
component of the unresolved emission.

We know that there are at least 4 distinct populations that could account  
for the unresolved emission: CVs, MSPs,
BY Dra and RS CVn binaries, and isolated
stellar coronae.
If we assume that the unresolved emission, about $3\times  
10^{32}$ \ergl (unabsorbed), is entirely due to stellar coronae and
there are about $10^5$ stars in the
volume that gives rise to the unresolved emission, then this implies that  
the average stellar corona radiates at $\sim 3\times10^{27}$ \ergl.
Since our Sun's X-ray luminosity varies between
$3\times10^{26}$ and $5\times10^{27}$ \ergl\ (Peres et al. 2000), we can  
say with confidence that the average star in M28 is less active than our  
Sun at its peak.
This is not unexpected, since X-ray activity appears to correlate with  
rotation, and the old stars in M28 are likely slowly rotating.
Therefore, we appeal to the usual suspects, faint CVs, MSPs, and RS CVn
and BR Dra binaries, to account for the excess background emission.
The X-ray luminosity functions for these classes are of course uncertain  
at the faint end.
Assuming average luminosities in the range $1 \times 10^{29}$ to $2 \times  
10^{30}$ \ergl implies 50--1000 such sources.

\subsection{Time Variability}\label{time_variability}

\begin{figure}
\centerline{\psfig{figure=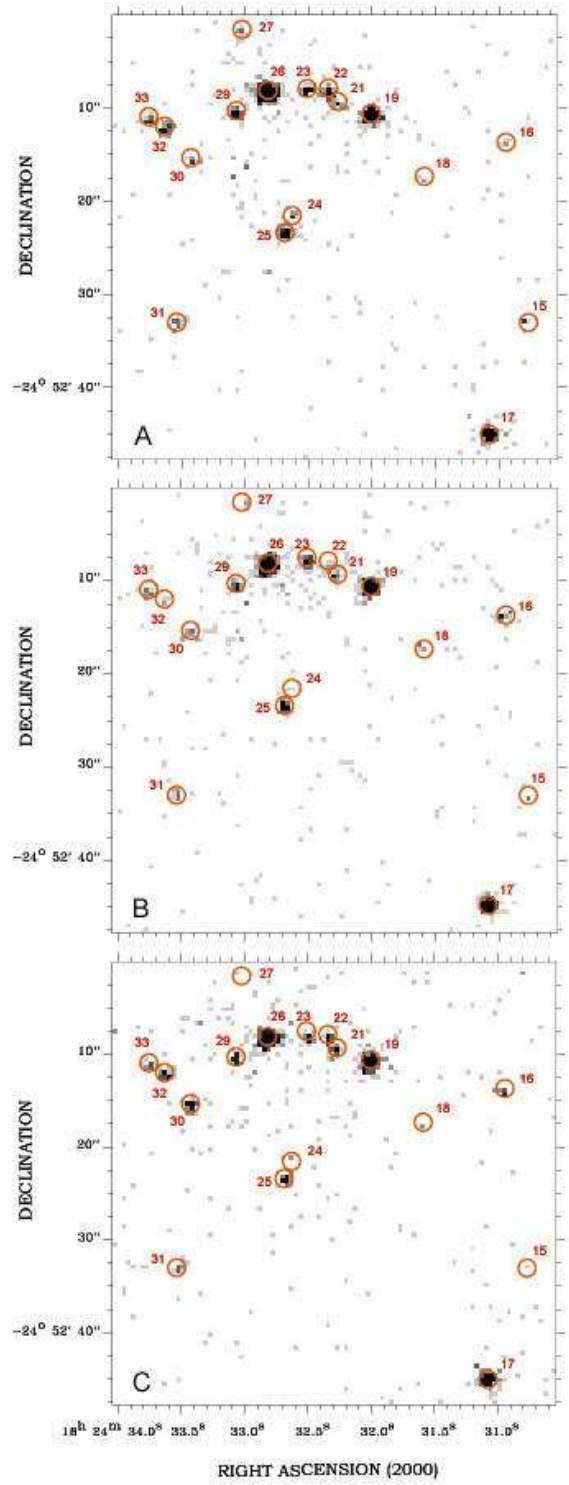,height=20cm,clip=}}
  \caption[]{  ACIS-S3 image of the central region of M28 at three
separate  
  epochs (A, B, C) as per Table~\ref{observations}. }
\label{M28abc}
\end{figure}

\begin{figure}
 \centerline{\psfig{figure=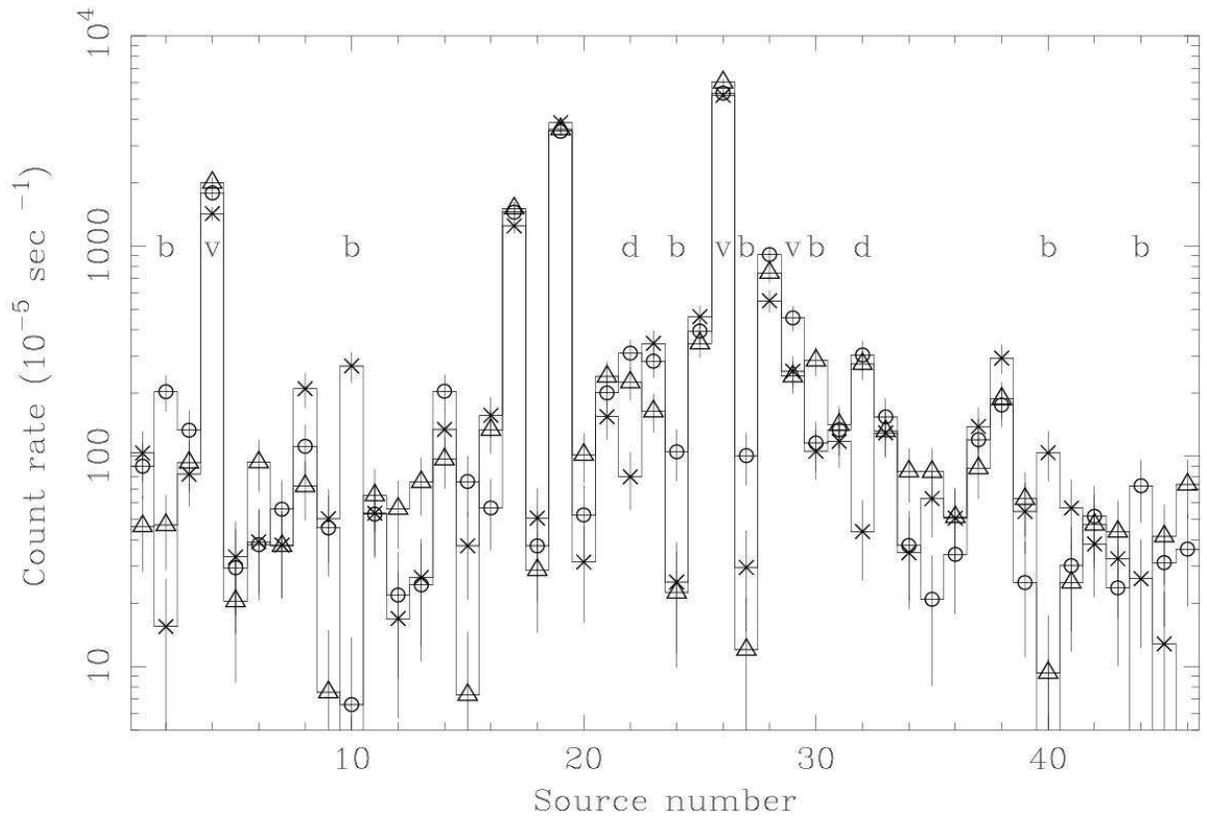,width=16cm,clip=}}
 \caption[]{
The count rate for all sources for each of the three observations.
The x-axis is the row number listed in Table~\ref{M28_source_table}.
For each source, three rates are plotted: a circle for the first  
observation, an ``x'' for the second, and a triangle for the third.
Any missing points were below the minimum rate plotted.
The statistical error bar is also plotted, but for the high count rate  
sources it is difficult to see.
The variability indicators (b, v and d) are discussed in  
\S~\ref{time_variability}.
}
\label{long_var}
\end{figure}

We have used the three available ACIS observations to search for source  
variability on a time scale of weeks.
The separate images are shown in Figure~\ref{M28abc}, while
Figure~\ref{long_var} shows the variation in the counting rates.
For all of the 46 sources, the number of counts detected in each of the 3  
observations was computed by modeling the spatial distribution of events  
with a two-dimensional circular gaussian function with a width increasing  
according to the off-axis angle to approximately match the point spread
function.
Contributions from nearby sources were accounted for by using the same  
gaussian function out to a distance of 7 $\sigma$ from the source  
considered.
In each case, both the position and width of the gaussians were kept  
constant in the data fitting, with only the normalizations left as free  
parameters.
This approach gives  a reasonable  estimate for the number of  counts from
each  source, even if the  source was  below the  detection threshold
and/or confused  with another nearby source.
As a check, the merged data was also examined and the results compared to  
the sums obtained from the 3 separate data sets.
The observed counting rates agreed to better than 1.5\%,
indicating no obvious biases in our procedure.

For each  source, we then calculated  the deviations of the number of
counts measured in each observation with respect to a constant flux
distribution.
By adding the 138 measurements ($46 \times3$), this  calculation yielded   
a $\chisq$ of 379.7 for 92 dof,
clearly showing that some of the sources actually varied.
The second observation of source \#10 contributed the most (44.7) to
$\chisq$, and so we designated that source as ``variable''.
The three observations of this source were then removed from the sample.
We then repeated this process.
In this way we determined that 12 of the 46 sources exhibit evidence for
some form of time variability.
Setting aside these 12 sources, $\chisq$ was 128.4 for 102 measurements and  
68 degrees of freedom.
The largest contribution now (\#28, 2nd measurement) contributes only
5.95 to $\chisq$ ($2.44\sigma$ deviation).
Although it is likely that this source also varies (as do a few others) the  
level is low enough to make characterizing the variability difficult, and  
thus we did not ascribe any designation of variability to these sources.

To characterize the variability further, if a source in one observation is  
more than a factor of 2 above the average for the other two measurements,  
we say that it brightened ("b" in Table~\ref{M28_source_table}).
Likewise, if one observation is more than a factor of 2 below the average  
for the other two measurements, we say that it dimmed ("d" in  
Table~\ref{M28_source_table}).
If the source neither ``brightened'' nor ``dimmed'' according to these  
definitions, but was still identified as varying we denote it with a "v".

The source identified as a possible qLMXB (\#26) was flagged as variable.
Since the emission is dominated by a very slowly cooling black body (\S2.2.2),
variability is not expected.
The variability designation is a consequence of the third observation  
being 13 percent ($4\sigma$) higher than the first two.
There are a number of possible systematic effects that might produce such  
an effect.
We verified the integration time for all three observations by counting  
the number of ACIS frames.
We also searched for missed  bad pixels and columns by checking which  
physical pixels were included in the extraction region, and found none.
Furthermore, the small change in the spacecraft roll (12 degrees) between  
observations combined with the spacecraft dither, resulted in a common set  
of pixels for all three measurements.
It is worth noting that cosmic ray tracks, the effects of which are  
discarded by the ACIS flight software, can produce up to 10 percent  
variations in
sensitivity, but of the front-illuminated (FI) chips.
Our observations, however, were performed with a back-illuminated CCD.
These are much thinner than the FI chips and consequently these tracks  
are about 10 times smaller, with a corresponding smaller impact on  
sensitivity.
In summary, we were unable to find any instrumental effect that could
account for the observed variability for source \#26.
On the other hand, the small amplitude of the variability suggests 
caution in applying the designation.

In addition to the millisecond pulsar, we were also able to compare
counting rates with one other ROSAT observation --- that of the source
with the fourth largest number of detected counts, \#17.
This source is located outside of the core radius at $42''$ from the
cluster's optical center. Although this source appears constant during 
our \CXO\ observations, {\sl ROSAT} observations indicate that the 
source is variable on time scales of  years. Specifically, this source 
was not detected in either the ROSAT PSPC observation of March 1991 nor 
the ROSAT HRI observation of September 1995. The source was, however, 
detected with ROSAT HRI in September 1996.

To summarize, in the \CXO\ data we found 6 sources that brightened, 3  
that were variable, and 2 that dimmed. In addition, we found one variable  
source from the ROSAT data.
Seven of these 13 sources are within one core radius of the center of
the globular cluster.
Of these seven, the three that brightened are most probably main sequence  
stars, whereas the two that were designated variable and the two that
dimmed are most likely CVs (or perhaps qLMXBs).
Finally, all the sources that appear to vary on shorter time scales 
(within one of the observations) had already been identified as varying 
by the technique described above.

\section{Optical observations \label{optical}}

  We have performed a search for potential  optical counterparts of the
  X-ray sources listed in Table  ~\ref{Opt_X_ident} using data obtained
  with the {\sl HST}  WFPC2 and available  in the public {\sl HST} data
  archive.   The observations  of  the M28 field  were  taken  with the
  ``V-band'' filter  F555W ($\lambda=   5500$  \AA; $\Delta  \lambda  =
  1200$\AA)  and  the  ``I-band''  filter F814W   ($\lambda=7995$  \AA;
  $\Delta  \lambda =  1292$ \AA) on  1997 September  12 (Testa et  al.\
  2001).  To allow for a better  cosmic ray filtering, the observations
  were split  into a sequence  of eight 140~s exposures in  the F555W
  filter, and three 180~s plus six 160~s exposures in the F814W filter.
  Three short exposures of 2.6~s each  were acquired in both filters to
  obtain unsaturated images   of  bright  cluster  stars.  The    total
  integration time   was  1130~s and  1510~s  in  the  F555W  and F814W
  filters, respectively. The same data have been used by Golden et 
  al.~(2001) to search for the optical counterpart of the MSP B1821-24.  
  
  Data reduction and photometric calibration were performed through the
  {\sl HST} WFPC2 pipeline.   For  each filter, single  exposures  were 
combined
using a cosmic ray  filter algorithm.  The final images
  were then registered on  each other.  Automatic object extraction and
  photometry  was    run by  using the  {\sl ROMAFOT}   package (Buonanno and
  Iannicola  1989).  The  source lists  derived for  each passband were
  finally matched to produce  the color catalog.  Conversion from pixel
  to sky  coordinates was computed  using  the task  {\sl metric} which
  also  applies the  correction  for the  WFPC2 geometrical distortions
  (see Testa et al.\ 2001 for further details on the data reduction and
  analysis). 
  
 The final catalog,  consisting   of a total   of 33972  entries,  with
 coordinates  and magnitudes in  the F555W passband and the F555W-F814W
 color, has  been used as a reference  for the  optical identification.
 As  a first step, the optical  catalog  has been cross-correlated with
 our list of X-ray sources.  Since the ACIS astrometry has been
 boresighted using the pulsar's  radio  coordinates as a reference  (\S
 \ref{image_analysis}),  the cross-correlation radius accounts only for
 the   statistical error of  the  X-ray  position  and for the
 uncertainty  of the {\sl HST} astrometry.   The latter  is ascribed to
 the  intrinsic error on the absolute  coordinates of  the GSC1.1 guide
 stars  used to   point the  {\sl   HST} and   compute the  astrometric
 solution.  Typical uncertainties  are of the order of 1\farcs0
 (see, e.g. Biretta et al.~ 2002)
  
\begin{figure}
  \centerline{\psfig{figure=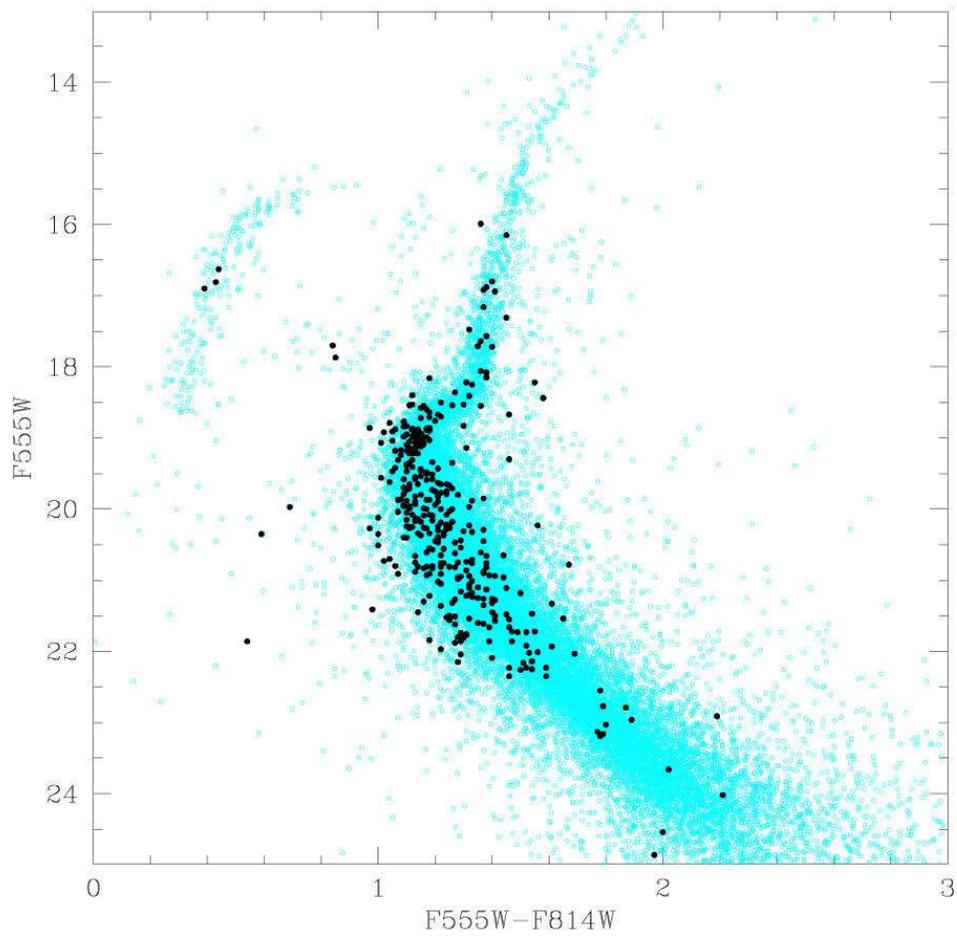,width=13cm,clip=}}
  \caption[]{
  Color-magnitude diagram for all the sources detected in the WFPC2 field of 
  view (light grey points). The magnitudes are corrected for the interstellar 
  extinction using the same E$(B-V)=0.43$ for all the sources. The 376 {\sl 
HST} 
  candidate counterparts matched to the X-ray sources listed in  
  Table~\ref{Opt_X_ident} are indicated by grey filled  circles. 
% Bigger star symbols mark the objects with large magnitude variation.
  }
 \label{hst_candidates_cmd1}
 \end{figure}

  Twenty-two  of the X-ray sources were  in  the   {\sl HST}
  field-of-view,  and  each of them   has potential WFPC2 counterparts.
  Figure~\ref{hst_candidates_cmd1} shows  the  location of all  the 376 
  matched WFPC2  sources on the  color-magnitude diagram  (CMD) derived
  from  all the sources   detected in the   WFPC2 field  of view.   The
  measured    magnitudes have  been    corrected for   the interstellar
  reddening assuming the  same  color excess E$(B-V)=0.43$   (Harris et
  al.~1996)  for all the sources.  Most  of  the candidate counterparts
  lie on the globular  cluster main sequence, with only  a few  of them
  possibly associated with evolved stellar  populations.  We note  that
  using any other average value of the reddening [e.g., E$(V-I)= 0.52$,
  derived directly  from our  datasets],  merely shifts the  whole CMD.
  However, a more  accurate analysis of the stellar
  populations   may require   accounting  for  possible   differential
  reddening along the line of sight (Testa et al.~2001). 
  This could narrow the distribution. 

  Additional F555W (340~s) and F814W (340~s) WFPC2 observations of the
  M28 field, taken on  1997  August 8, have  been  used to search   for
  variability among  the potential WFPC2  counterparts.   The data have
  been retrieved from the ST-ECF public archive\footnote{www.stecf.org}
  after  on-the-fly recalibration  with  the best  available  reference
  files.   The two pointings are centered  very close to each other but
  with  a slight relative  rotation angle ($\sim \mathrm{5}^{\circ}$ ).
  Image co-addition,  object detection  and position  measurements were
  performed consistent with the previous analyses. 
 To avoid  systematic effects  due to the   difference  in the  default
  astrometric  solution   between  the two  datasets,   which  has been
  measured  to  be of the   order of 1\farcs0, the  object catalogs
  derived from the August 8 and  September 12 observations were matched
  in  the pixel space  after  registering the  two sets  of coordinates
  through   a linear transformation.   The  overall  dispersion of  the
  radial coordinate residuals after the transformation turned out to be
  0.17 WFC pixels (0\farcs017). For this reason, only objects with radial
  coordinate residuals smaller than 0.5  WFC pixels, i.e. 3 $\sigma$
  of the dispersion of the residuals, 
  were considered  as
  matched in the two datasets and examined for variability.  Possible
  spurious matches were checked manually and filtered out. 

 Although a number of candidates  show brightness variation larger than
 0.5 magnitudes in at least one of the  filters, their nature cannot be
 assessed with  high confidence. Some of  them are indeed close  to the
 detection limit, hence with larger photometry errors, while others are
 detected too close  to bright stars  to obtain clean measurements even
 using PSF  subtraction  alghoritms, and  a few are located in very
 crowded patches. 

 We  note that some objects detected in the first dataset are absent in the
second. All  these  cases have been   checked
 carefully to  find out whether the lack  of matches  was due to 
 intrinsic  object variability.  However, we found that the missing matches 
 can be explained either by the fact that in the second dataset some objects 
 fall at the chip edge or in the overscan region, or they fall out of  the 
 field of view or, as the second dataset is shallower, they are simply 
 below the detection limit. 

  Although we have  found  a number of  candidate counterparts  to  the
  X-ray sources,   no definite conclusions can   yet be drawn  from our
  search.  In   the crowded   globular   cluster field,   the  dominant
  uncertainty   in  the  {\sl   HST} astrometry   represents   the real
  bottleneck for obtaining optical identifications.  In contrast, e.g.,
  to the experience of  Pooley et al.\  (2002a) with NGC 6752, we found
  very  few  blue  candidates  which   can  be  considered  as   likely
  counterparts and used to boresight the  {\sl HST} astrometry.  At the
  same time, no GSC2.1  or USNO-A2.0  stars are  present in the  narrow
  WFPC2 field of view to use them as a reference to recompute the image
  astrometric  calibration.   The  only  way    to  improve the   WFPC2
  astrometry  of our datasets  is by upgrading  the coordinates and the
  positional  accuracy  of the  guide  stars   used for  the  telescope
  pointing and recomputing the  astrometric  solution in the  {\sl HST}
  focal plane.  This, together with   a larger spectral coverage,  will
  definitely give us a better chance to obtain firm identifications.

\section{Summary}

We have analyzed observations of the globular cluster M28
taken with the ACIS-S3 instrument aboard the \CXO.
Forty-six X-ray sources were detected within $3\farcm 1$ of the optical  
center of the cluster down to a limiting (absorption-corrected) luminosity  
of $\sim 6.8 \times 10^{30}$~\ergl\ in the 0.5~--~8.0~keV range.
Many of these sources are concentrated near the center of the cluster.
Their radial distribution can be described by a King profile with an X-ray 
core radius comparable to that of optical light but with a steep power-law  
index suggesting a rather high X-ray source population mass, $M_x \sim  
1.9$~\msun. The X-ray source distribution flattens at larger radii consistent 
with a population of background sources.

Among the brightest sources in M28 is the millisecond pulsar \psr.
We find the phase-averaged spectrum of \psr\ to be best represented by
a $\Gamma=1.20$ power law radiating at $L_x = 1.3\times 10^{33}$~\ergl.
The luminosity is consistent with a steady luminosity since the time of
\ROSAT\ observations.

An intriguing spectral feature, albeit of marginal statistical significance,
observed at $\sim$3~keV in \psr\ might be an electron cyclotron line.
If so, the line requires a magnetic field of order 100 times the strength
inferred from $P$ and $\dot{{P}}$ suggesting the local magnetic field at
the surface of the neutron star is well above the conventional values
obtained by assuming a centered dipole geometry. This may be due to  
multipolar components, to a strong off-centering of the magnetic dipole,
or to a combination of the two. Recently, Gil \& Melikidze (2002)
argued on both observational and theoretical grounds that such
strong deviations from the dipole magnetic field should exist at or near 
pulsar
polar caps.
Geppert \& Rheinhardt (2002) showed that it is possible to create strong but
small-scale poloidal field structures at the neutron star surface via a
Hall-instability from subsurface
toroidal field components.

\begin{figure}
 \centerline{\psfig{figure=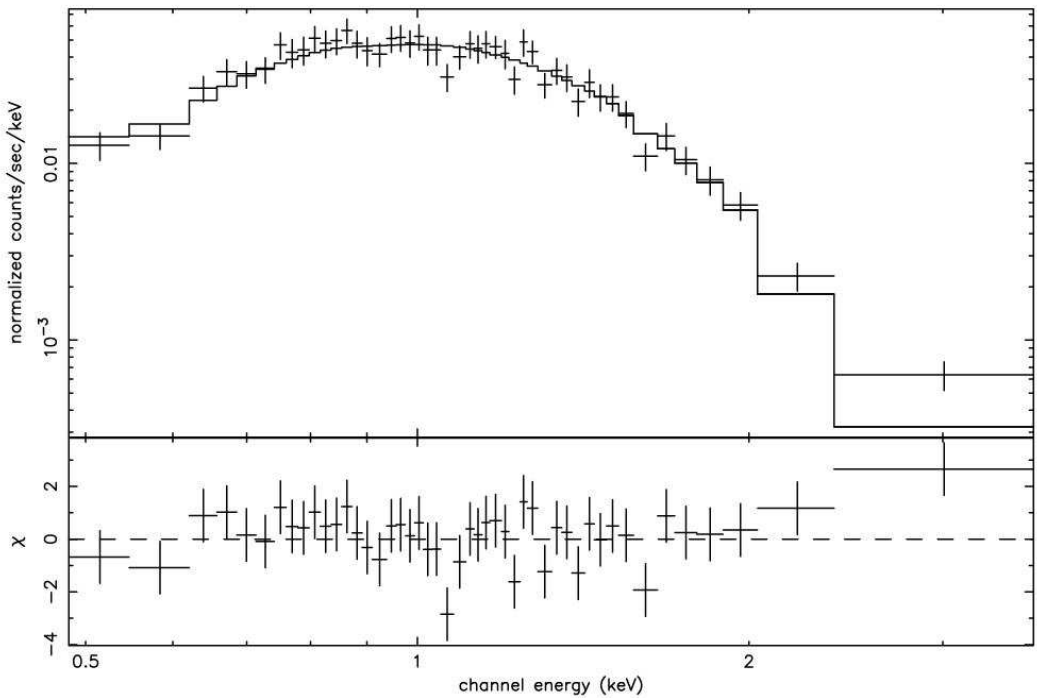,width=14cm,clip=}}
 \caption[]{Energy spectrum of source \#26. The spectrum is modeled with a
nonmagnetic 
 neutron star H atmosphere model.}
 \label{bright_26}
\end{figure}

A second bright source, closer to the center of M28, was also
studied in detail. The spectrum of this source (\#26) is notably soft and
thermal (Fig.~\ref{bright_26}) as is typical of transiently-accreting
neutron stars in quiescence. Such objects have a hydrogen-rich atmosphere
comprised of matter accumulated and heated during previous accretion
episodes. Non-magnetic hydrogen atmosphere models
provide the best fit to the spectrum of this source and thereby
support this interpretation. The bolometric luminosity corresponding 
to the best fit model parameters, $L_{\rm bol}^{\infty} \sim
1.9 \times 10^{33}$~\ergl, can be maintained by a time-averaged 
accretion rate of $\sim 2\times 10^{-11}$~\msun~yr$^{-1}$.
While the flux from this source varies among the 3 \CXO\ observations,
there is no evidence for spectral variability between our observations
taken at roughly 1 month intervals.

While \CXO\ resolves many of the X-ray sources in M28, there remains some
diffuse emission distributed over $\sim$1 core radius. This
emission is only 22\% of the total X-ray luminosity in the central region.
Nevertheless, simple extrapolation of the observed $\log N$~--~$\log S$
relation to below our detection threshold cannot account for more than
about one-half of the emission. Another population of weak
sources must account for the remainder. Possibilities include
BY~Dra and (weak) RS~CVn systems, other millisecond pulsars, isolated
low-mass stars, and CVs.

By scheduling our 3 observations of M28 over an $\sim$2 month period,
we were able to observe variability in the resolved X-ray source
population on a time scale of order weeks.
Twelve of the sources, or 26\%, are seen to vary over the course of the
observations including the quiescent low-mass X-ray binary. Some sources
exhibited a much higher flux in 1 of the 3 observations, and may be associated
with stellar coronae. Others display the opposite effect, being low in one  
observation or otherwise varied and may be associated with CVs and qLMXBs.

Finally, the benefit of an accurate radio position for \psr\ has allowed us to
constrain the X-ray positions of the M28 sources to sub-pixel accuracy. 
Comparison to \HST\ images suffers, however, from the intrinsic error in 
the absolute coordinates of the GSC1.1 guide stars used to compute the WFPC2 
astrometric solution thus preventing any conclusive identifications.

\acknowledgments
Those of us at the Marshall Space Flight Center acknowledge support from the
{\sl Chandra} Project.
G.G.P. acknowledges support from NASA grant NAG5-10865. V.T. acknowledges
support from MIUR 
under grant COFIN 2001: "Modeling old Millisecond Pulsars in the Galaxy and in
Globular Clusters"

\clearpage

\clearpage

\clearpage

\begin{figure}
\centerline{\psfig{figure=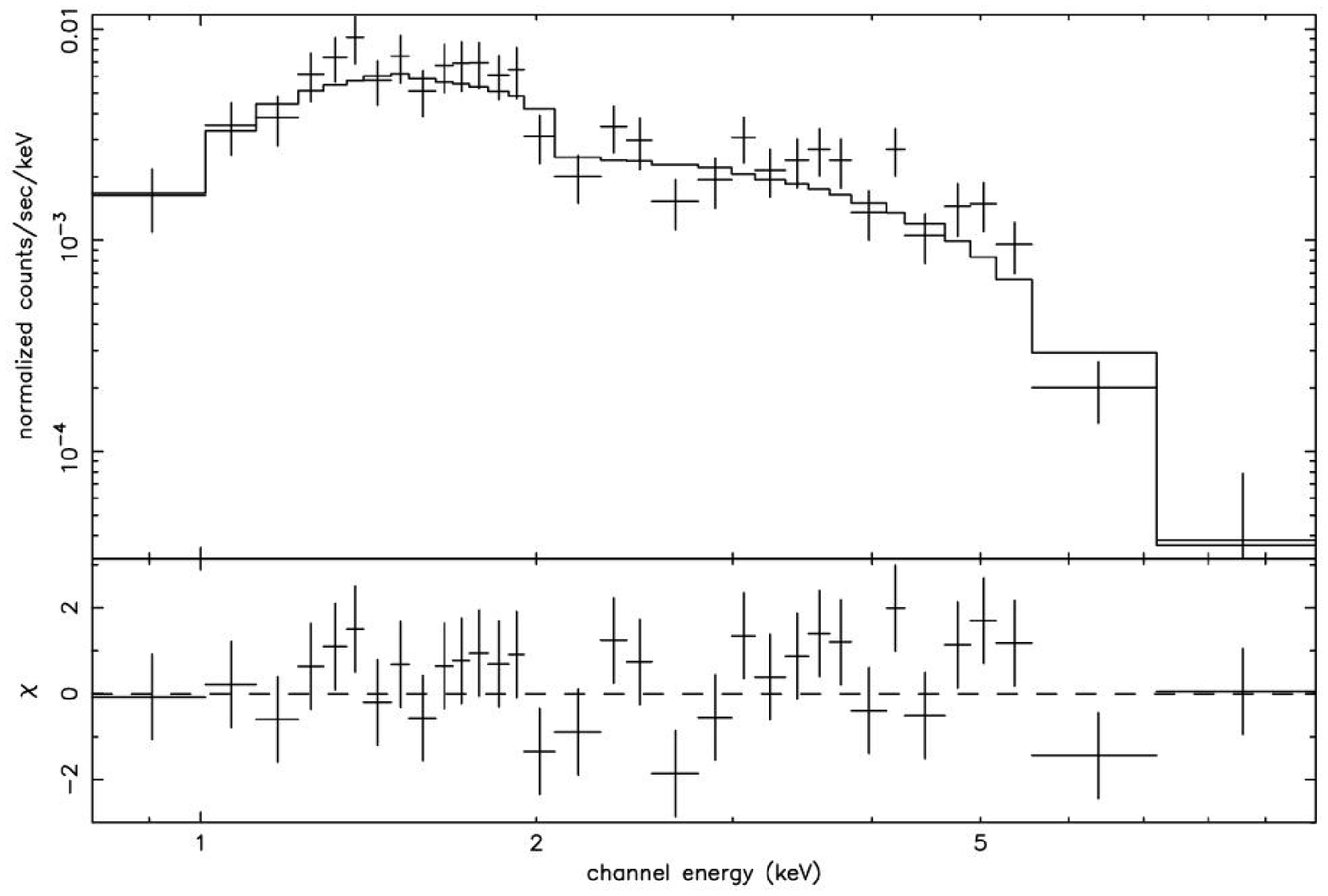,width=14cm,clip=}}
 \caption[]{Energy spectrum of source \#4, fit with an absorbed power-law
model.}
 \label{bright_source_4_spectrum}
\end{figure}

\clearpage

\clearpage

\clearpage

\clearpage

 \begin{deluxetable}{ccc}
  \tablewidth{0pc}
  \tablecaption{ \CXO\ Observations \label{observations}}
  \tablehead{}
  \startdata
   Date         &  ObsID   &  Exposure       \\[1ex]\hline\\[-2ex]
   2002\, July 4  &   2684   &     12.7 ksec   \\
   2002\, Aug  8  &   2685   &     13.5 ksec   \\
   2002\, Sep  9   &  2683   &     11.4 ksec   \\\hline
  \enddata
 \end{deluxetable}

 \begin{deluxetable}{cccc}
  \tablewidth{0pc}
  \tablecaption{ PSR B1821$-$24 Positions (J2000)\label{psr_positions}}
  \tablehead{}
  \startdata
   Date & Position \\ \hline\\[-1.5ex]
   2002 July 4 &  18 24 32.015 &  $-$24 52 10.81 \\
   2002 Aug  8 &  18 24 32.016 &  $-$24 52 10.76 \\
   2002 Sep  9 &  18 24 32.009 &  $-$24 52 10.83 \\
   average     &  18 24 32.013 &  $-$24 52 10.80 \\
   rms (arcsec)& 0.042         &  0.029        \\
   merged data & 18 24 32.013  & $-$24 52 10.80  \\
   radio (8/02)& 18 24 32.008  & $-$24 52 10.76  \\\hline\\[-1.5ex]
 \enddata
 \end{deluxetable}

\clearpage
\begin{deluxetable}{rrrrrrrrrrc} \label{Opt_X_ident}
%\rotate
\tabletypesize{\scriptsize}
   \tablewidth{0pc}
   \tablecaption{M28 Discrete X-Ray Sources\label{M28_source_table} \label{Opt_X_ident}}
    \tablehead{}
   \startdata
%
%\hline \hline
%
 &  \multicolumn{1}{c}{R.A.} & \multicolumn{1}{c}{Dec.} & \multicolumn{1}{c}{$r^a$} & \multicolumn{1}{c}{$d^b$} & \multicolumn{1}{c}{S/N} &
\multicolumn{1}{c}{Soft$^c$} & \multicolumn{1}{c}{Medium$^d$} & \multicolumn{1}{c}{Hard$^e$} & \multicolumn{1}{c}{$L_X{^f}$} & Variability$^g$ \\ &  \multicolumn{1}{c}{(J2000)} & \multicolumn{1}{c}{(J2000)}
 & ($\arcsec$) &($\arcsec$) & &($10^{-5}$~s$^{-1}$) & ($10^{-5}$~s$^{-1}$) & ($10^{-5}$~s$^{-1}$) & \multicolumn{1}{c}{($10^{30}$~erg~s$^{-1}$)} & \\ \hline
  1$\,\,\,\,\,\,\,$ & 18 24 20.531 &  -24 51 33.04 &  0.433 & 172 &  4.09 & $  15.7_{-7.7}^{+7.0}  $ & $26.1_{-9.2}^{+8.6}    $ & $19.1_{-8.2}^{+7.6}    $  & $ 17.0$ &  \\
  2$\,\,\,\,\,\,\,$ & 18 24 20.619 &  -24 51 27.17 &  0.431 & 172 &  4.16 & $ 15.8_{-7.8}^{+7.1}  $ & $35.3_{-10.6}^{+10.0}  $ & $13.3_{-7.2}^{+6.5}    $  & $ 17.8$ & b\\
  3$\,\,\,\,\,\,\,$ & 18 24 22.575 &  -24 52 05.69 &  0.361 & 140 &  5.02 & $ 35.5_{-10.8}^{+10.2}$ & $36.3_{-10.8}^{+10.2}  $ & $17.3_{-7.9}^{+7.2}    $  & $ 25.1$ &  \\
  4$\,\,\,\,\,\,\,$ & 18 24 22.684 &  -24 51 02.65 &  0.303 & 154 & 21.12 & $ 47.9_{-12.2}^{+11.7} $ & $663.4_{-43.3}^{+42.8} $ & $756.5_{-46.3}^{+45.8}  $ & $578.9$ & v\\
  5$\,\,\,\,\,\,\,$ & 18 24 22.831 &  -24 52 46.06 &  0.469 & 141 &  2.60 & $ 13.2_{-6.8}^{+6.2}  $ & $8.2_{-10.6}^{+4.8}    $ & $2._{-2.0}^{+8.7}      $  & $  6.5$ &  \\
  6$\,\,\,\,\,\,\,$ & 18 24 24.237 &  -24 51 04.14 &  0.421 & 135 &  3.75 & $ 14.9_{-8.2}^{+7.4}  $ & $36.8_{-11.6}^{+10.9}  $ & $1.7_{-1.7}^{+8.8}     $  & $ 15.3$ &  \\
  7$\,\,\,\,\,\,\,$ & 18 24 24.603 &  -24 53 02.10 &  0.392 & 123 &  3.14 & $  20.1_{-9.1}^{+8.3}  $ & $17.3_{-8.5}^{+7.6}    $ & $2.7_{-2.7}^{+11.1}    $  & $  9.9$ &  \\
  8$\,\,\,\,\,\,\,$ & 18 24 25.175 &  -24 54 06.67 &  0.327 & 155 &  5.75 & $ 
54.8_{-14.1}^{+13.4} $ & $55.5_{-14.1}^{+13.4}  $ & $27.1_{-10.3}^{+9.5}    $ 
&
$ 33.4$ &  \\
  9 *& 18 24 25.189 &  -24 52 13.60 &  0.408 & 104 &  2.82 & $  
0_{-0.0}^{+4.2}    
$ & $0_{-0.0}^{+4.2}       $ & $36.5_{-12.2}^{+11.4}  $  & $  9.3$ &  \\
 10$\,\,\,\,\,\,\,$ & 18 24 25.658 &  -24 50 34.14 &  0.407 & 138 &  4.58 & $ 
21.3_{-8.6}^{+7.9}  $ & $31.3_{-9.9}^{+9.3}    $ & $12.9_{-7.0}^{+6.3}    $  &
$ 20.4$ & b\\
 11$\,\,\,\,\,\,\,$ & 18 24 28.449 &  -24 50 33.59 &  0.423 & 114 &  3.72 & $ 
11.4_{-7.4}^{+6.5}  $ & $32.9_{-10.9}^{+10.2}  $ & $11.8_{-7.4}^{+6.5}    $  &
$ 14.6$ &  \\
 12 *& 18 24 28.580 &  -24 53 09.77 &  0.359 &  82 &  3.11 & $  
0_{-0.0}^{+3.2}    
$ & $17.1_{-7.5}^{+6.8}    $ & $16.9_{-7.5}^{+6.8}    $  & $  9.7$ &  \\
 13 *& 18 24 28.727 &  -24 51 24.56 &  0.405 &  73 &  3.23 & $ 
10._{-6.1}^{+5.3}   $ & $13.4_{-6.6}^{+6.0}    $ & $10.2_{-6.1}^{+5.3}    $  &
$ 10.2$ &  \\
 14 *& 18 24 30.155 &  -24 51 49.81 &  0.323 &  42 &  5.76 & $ 
18.9_{-8.5}^{+7.8}  $ & $68.7_{-15.1}^{+14.5}  $ & $48.7_{-12.9}^{+12.2}  $  &
$ 34.8$ &  \\
 15 *& 18 24 30.770 &  -24 52 33.19 &  0.367 &  36 &  2.91 & $ 
13.7_{-6.9}^{+6.2}  $ & $14.1_{-6.9}^{+6.2}    $ & $0_{-0.0}^{+3.2}       $  &
$  8.3$ &  \\
 16 *& 18 24 30.946 &  -24 52 13.84 &  0.323 &  26 &  5.12 & $ 
23.3_{-8.8}^{+8.1}  $ & $53.5_{-12.8}^{+12.2}  $ & $32.3_{-10.1}^{+9.5}   $  &
$ 27.5$ &  \\
 17 *& 18 24 31.063 &  -24 52 45.20 &  0.299 &  41 & 18.77 & $
254.3_{-26.3}^{+25.8}$ & $537.6_{-38.1}^{+37.6} $ & $514.9_{-37.3}^{+36.8}  $ 
&
$439.1$ & v$^h$\\
 18 *& 18 24 31.591 &  -24 52 17.49 &  0.377 &  18 &  2.88 & $ 
5.3_{-5.3}^{+19.9}  $ & $14.5_{-7.0}^{+6.3}    $ & $11.3_{-6.5}^{+5.7}    $  &
$  7.8$ &  \\
 19 *& 18 24 32.008 &  -24 52 10.76 &  0.298 &  11 & 30.09 & $
526.4_{-38.9}^{+38.4}$ & $1369_{-63}^{+62}      $ & $1405_{-63}^{+63}       $ 
&
$ 96.0$ &  \\
 20$\,\,\,\,\,\,\,$ & 18 24 32.213 &  -24 53 51.58 &  0.331 & 100 &  3.69 & $ 
21.5_{-8.0}^{+7.5}  $ & $24.5_{-8.5}^{+7.9}    $ & $13.4_{-6.6}^{+5.9}    $  &
$ 15.6$ &  \\
 21 * & 18 24 32.272 &  -24 52 09.46 &  0.323 &   8 &  4.85 &$ 
29.5_{-9.7}^{+9.1}  $ & $77.6_{-15.2}^{+14.7}  $ & $47.5_{-12.1}^{+11.5}  $  &
$ 27.6$ &  \\
 22 * & 18 24 32.345 &  -24 52 08.02 &  0.316 &   8 &  5.30
&$35.4_{-10.6}^{+10.0} $ & $116.4_{-18.5}^{+18.0} $ & $68.4_{-14.4}^{+13.8}   
$
& $ 37.2$ & d\\
 23 * & 18 24 32.514 &  -24 52 07.66 &  0.312 &   6 &  6.73
&$44.3_{-11.7}^{+11.2} $ & $95.6_{-16.9}^{+16.3}  $ & $137.3_{-20.1}^{+19.6}  
$
& $ 64.5$ &  \\
 24 * & 18 24 32.631 &  -24 52 21.70 &  0.354 &  10 &  3.35 & $ 
13.7_{-6.9}^{+6.2}  $ & $22.7_{-8.4}^{+7.8}    $ & $13.8_{-6.9}^{+6.2}    $  &
$ 12.0$ & b\\
 25 * & 18 24 32.689 &  -24 52 23.54 &  0.304 &  12 &  9.65 & $ 
54._{-12.6}^{+12.1} $ & $129.1_{-19.1}^{+18.6} $ & $172._{-22.0}^{+21.5}  $  &
$129.0$ &  \\
 26 * & 18 24 32.821 &  -24 52 08.26 &  0.298 &   3 & 36.93 & 
$2213.8_{-79.5}^{+79.}$ & $2477.1_{-84.1}^{+83.6}$ & $211.7_{-24.9}^{+24.3}  $
& $534.0$ & v\\
 27 * & 18 24 33.026 &  -24 52 01.72 &  0.353 &   9 &  3.46 & $ 
8.7_{-11.3}^{+5.1}  $ & $12._{-6.7}^{+5.9}     $ & $24._{-8.9}^{+8.3}     $  &
$ 12.7$ & b\\
 28$\,\,\,\,\,\,\,$ & 18 24 33.026 &  -24 50 52.86 &  0.306 &  78 & 13.72 & 
$42.9_{-11.5}^{+10.9} $ & $278._{-28.1}^{+27.6}  $ & $298.1_{-29.1}^{+28.6}  $
& $207.9$ &  \\
 29 *& 18 24 33.070 &  -24 52 10.45 &  0.308 &   2 &  7.90 & 
$67.6_{-14.3}^{+13.7} $ & $124.4_{-19.1}^{+18.5} $ & $67.8_{-14.3}^{+13.7}   $
& $ 65.8$ & v\\
 30 *& 18 24 33.429 &  -24 52 15.49 &  0.313 &   8 &  6.50 & 
$34.6_{-10.3}^{+9.8}  $ & $76._{-14.9}^{+14.4}   $ & $52.3_{-12.5}^{+12.0}   $
& $ 42.8$ & b\\
 31$\,\,\,\,\,\,\,$ & 18 24 33.539 &  -24 52 33.16 &  0.316 &  23 &  5.62 & $ 
13.6_{-6.8}^{+6.1}  $ & $61.9_{-13.3}^{+12.7}  $ & $36.3_{-10.3}^{+9.8}   $  &
$ 31.5$ &  \\
 32 *& 18 24 33.634 &  -24 52 12.12 &  0.312 &  10 &  6.74 & $ 
17.2_{-7.7}^{+7.0}  $ & $103.4_{-17.4}^{+16.9} $ & $67.6_{-14.2}^{+13.7}  $  &
$ 47.3$ & d\\
 33 *& 18 24 33.759 &  -24 52 11.08 &  0.321 &  11 &  4.87 & 
$38.1_{-10.9}^{+10.3} $ & $71.1_{-14.6}^{+14.0}  $ & $20.4_{-8.2}^{+7.6}     $
& $ 29.4$ &  \\
 34 *& 18 24 33.861 &  -24 51 11.99 &  0.391 &  60 &  3.59 & $ 
7.4_{-10.2}^{+4.6}  $ & $24.5_{-8.5}^{+8.0}    $ & $7.5_{-10.2}^{+4.6}    $  &
$ 11.8$ &  \\
 35 *& 18 24 34.469 &  -24 53 12.99 &  0.335 &  65 &  3.67 & $ 
26.9_{-8.8}^{+8.3}  $ & $21.7_{-8.0}^{+7.4}    $ & $0_{-0.0}^{+3.}        $  &
$ 13.4$ &  \\
 36$\,\,\,\,\,\,\,$ & 18 24 34.974 &  -24 54 57.88 &  0.368 & 168 &  3.39 & $ 
18.5_{-8.3}^{+7.6}  $ & $28.7_{-10.}^{+9.3}    $ & $12.2_{-7.1}^{+6.2}    $  &
$ 11.3$ &  \\
 37$\,\,\,\,\,\,\,$ & 18 24 35.524 &  -24 52 18.34 &  0.321 &  36 &  5.45 & $ 
14.3_{-7.1}^{+6.4}  $ & $53.3_{-12.7}^{+12.1}  $ & $35.2_{-10.5}^{+9.9}   $  &
$ 29.5$ &  \\
 38$\,\,\,\,\,\,\,$ & 18 24 36.578 &  -24 50 16.04 &  0.340 & 125 &  7.58 & 
$52.9_{-12.6}^{+12.1} $ & $74.3_{-14.7}^{+14.2}  $ & $64.8_{-13.9}^{+13.3}   $
& $ 56.5$ &  \\
 39$\,\,\,\,\,\,\,$ & 18 24 37.326 &  -24 51 57.50 &  0.363 &  61 &  3.46 & $ 
25.9_{-9.8}^{+9.1}  $ & $16.4_{-8.0}^{+7.2}    $ & $2.7_{-2.7}^{+10.8}    $  &
$ 13.2$ &  \\
 40$\,\,\,\,\,\,\,$ & 18 24 37.831 &  -24 51 44.58 &  0.401 &  72 &  2.93 & $ 
13.1_{-7.9}^{+6.9}  $ & $17.5_{-8.6}^{+7.7}    $ & $2.6_{-2.6}^{+11.1}    $  &
$  9.1$ & b\\
 41$\,\,\,\,\,\,\,$ & 18 24 37.897 &  -24 49 28.45 &  0.632 & 176 &  2.75 & $ 
2.5_{-2.5}^{+10.7}  $ & $19.4_{-8.1}^{+7.5}    $ & $9.1_{-6.4}^{+5.7}     $  &
$  8.6$ &  \\
 42$\,\,\,\,\,\,\,$ & 18 24 39.061 &  -24 51 08.26 &  0.434 & 105 &  3.15 & $ 
4.7_{-4.7}^{+18.}   $ & $2.5_{-2.5}^{+9.9}     $ & $29.1_{-9.8}^{+9.2}    $  &
$ 10.3$ &  \\
 43$\,\,\,\,\,\,\,$ & 18 24 39.567 &  -24 50 30.18 &  0.561 & 136 &  2.63 & $ 
12.5_{-6.9}^{+6.3}  $ & $10.9_{-6.4}^{+5.6}    $ & $1.3_{-1.3}^{+7.3}     $  &
$  6.8$ &  \\
 44$\,\,\,\,\,\,\,$ & 18 24 40.336 &  -24 50 35.31 &  0.563 & 139 &  2.64 & $ 
12.5_{-7.}^{+6.3}   $ & $5.1_{-5.1}^{+19.4}    $ & $4.2_{-4.2}^{+16.3}    $  &
$  6.8$ & b\\
 45$\,\,\,\,\,\,\,$ & 18 24 41.303 &  -24 54 16.97 &  0.419 & 170 &  2.63 & $ 
0_{-0.0}^{+4.1}     $ & $25.4_{-10.2}^{+9.4}   $ & $2.7_{-2.7}^{+11.4}    $  &
$  8.1$ &  \\
 46$\,\,\,\,\,\,\,$ & 18 24 42.694 &  -24 52 47.48 &  0.433 & 138 &  2.81 & $ 
10.1_{-6.2}^{+5.4}  $ & $13.6_{-6.7}^{+6.0}    $ & $4.7_{-4.7}^{+18.2}    $  &
$  8.1$ &  \\
\hline
\enddata
 \tablecomments{\\
\\
Sources indicated with a * are in the field of view of the {\sl HST}  
observations discussed in \S~\ref{optical} \\
$^a$ Positional uncertainty radius in arcsec (see text).\\
$^b$ Distance from the nominal optical center of the cluster.\\
$^c$ Detected counting rate, corrected for the psf, in the 0.2--1.0 keV 
band.\\
$^d$ Detected counting rate, corrected for the psf, in the 1.0--2.0 keV 
band.\\
$^e$ Detected counting rate, corrected for the psf, in the 2.0--8.0 keV 
band.\\
$^f$ X-ray luminosity in the 0.5--8.0 keV band assuming a distance of 5.5
kpc and $N_H$ of $0.18 \times 10^{22}$ cm$^{-2}$. The luminosities of the
six brightest sources are based on a canonical power-law. More accurate
luminosities for the six brightest sources are presented in the text. \\
$^g$ v indicates variability, f a flare and d a dip as discussed in
\S~\ref{time_variability}.\\
$^h$ based on comparison with ROSAT observations.
}
 \end{deluxetable}

\clearpage
\begin{deluxetable}{ccccccc}
  \tablewidth{0pc}
  \tablecaption{Extraction and spectral fitting parameters\label{bright_5_t1}}
  \tablehead{}
  \startdata
source \# &     $r_{\rm ext}^a$ & f$^b$ & $N_{S+B}^c$ & $N_{B}^d$ & $N_{\rm
min}^e$ & $N_{\rm bins_{_{}}}^f$\\ \hline\\[-1.5ex]
19 & $1\farcs72$  & 97.5 & 1119 &  2   & 30 & 34 \\
26 & $1\farcs72$  & 97.5 & 1669 &  2   & 30 & 47 \\
4  & $8\farcs9$   & 100  & 540  & 49   & 15 & 33 \\
17 & $7\farcs5$   & 100  & 527  & 28   & 20 & 24 \\
28 & $14\farcs8$  & 95   & 300  & 115  & 15 & 18 \\
25 & $1\farcs35$  & 95   & 127  & 1    & 10 & 11 \\ \hline

 \enddata
 \tablecomments{\\
\\
$^a$ Extraction radius\\
$^b$ Approximate percentage of total signal\\
$^c$ Total number of extracted counts.\\
$^d$ Estimated number of background counts.\\
$^e$ Minimum number of counts per spectral bin. \\
$^f$ Number of spectral bins.\\
}

 \end{deluxetable}
\clearpage
\begin{deluxetable}{ccccccccc}
  \tablewidth{0pc}
  \tablecaption{Spectral Fit
  \label{bright_5_t2}}
  \tablehead{}
  \startdata
  source & model$^a$ & $\chi_\nu^2$ & $\nu$ &      $N_H/10^{22}$      &
$\Gamma$ or $kT$$^b$      &        Radius$^c$        & 
$\mbox{Flux}^d/10^{-13}$               
\\
  number &  {}       &     {}       &   {}  &   $\mbox{cm}^{-2}$      &         
{}              &          km              &         \fcgs            
\\\hline\\[-1ex]
   19 (PSR)   &  pl       &     0.89     &  31   & $0.16^{+0.07}_{-0.08}$ & 
$1.20^{+0.15}_{-0.13}$  &          {}              & $3.54^{+0.06}_{-0.05}$   
\\\\[-1ex]\hline\\[-1ex]
   26    &  bb       &     1.10     &  44   & $0.13_{-0.05}^{+0.05}$  &
$0.26_{-0.18}^{+0.18}$  & $1.27_{-0.23}^{+0.29}$   & $1.59_{-0.71}^{+1.38}$   
\\\\[-1ex]
   26    &  pl       &     0.86     &  44   & $0.68_{-0.07}^{+0.08}$  &
$5.24^{+0.39}_{-0.35}$  &          {}              & $14.8_{-3.95}^{+0.65}$   
\\\\[-1ex]
   26    &  nsa      &     0.96     &  44   & $0.26_{-0.04}^{+0.04}$  &
$0.09^{+0.03}_{-0.01}$  & $14.5_{-3.8}^{+6.9}$     & $3.35_{-1.1}^{+1.9}$     
\\\\[-1ex]
   26    & mekal     &     0.88     &  44   & $0.33_{-0.05}^{+0.02}$  &
$0.59^{+0.06}_{-0.06}$  &          {}              & $3.26_{-1.04}^{+1.60}$   
\\\\[-1ex]\hline\\[-1ex]
    4    &  bb       &     1.39     &  30   & $0.16_{-0.13}^{+0.16}$  &
$1.07^{+0.13}_{-0.11}$  &$0.063_{-0.01}^{+0.02}$   & $1.69_{-0.95}^{+0.66}$   
\\\\[-1ex]
    4    &  pl       &     1.14     &  30   & $0.86_{-0.18}^{+0.26}$  &
$1.59^{+0.15}_{-0.24}$  &          {}              & $2.49_{-0.26}^{+0.71}$   
\\\\[-1ex]
    4    &  mekal    &     1.14     &  30   & $0.77_{-0.17}^{+0.19}$  &
$13.3^{+41.4}_{-6.7}$   &          {}              & $2.34_{-0.48}^{+0.14}$   
\\\\[-1ex]\hline\\[-1ex]
   17    &  bb       &     1.73     &  21   & $0.0-0.028$             &
$0.88^{+0.08}_{-0.76}$  & $0.073_{-0.01}^{+0.01}$  & $1.07^{+0.89}_{-0.49}$   
\\\\[-1ex]
   17    &  pl       &     1.02     &  21   & $0.17^{+0.15}_{-0.1}$   &
$1.30^{+0.21}_{-0.18}$  &          {}              & $1.47^{+0.15}_{-0.11}$   
\\\\[-1ex]
   17    & mekal     &     0.99     &  21   & $0.17^{+0.11}_{-0.17}$  &
$38^{+42}_{-26}$        &          {}              & $1.44^{+0.13}_{-0.19}$   
\\\\[-1ex]\hline\\[-1ex]
   28    &  bb       &     0.67     &  15   & $0.64^{+0.52}_{-0.36}$  &
$0.67^{+0.14}_{-0.12}$  &  $0.095_{-0.03}^{+0.06}$ & $0.59^{+2.76}_{-0.48}$    
\\\\[-1ex]
   28    &  pl       &     0.74     &  15   & $1.83^{+0.93}_{-0.63}$  &
$3.08^{+1.00}_{-0.67}$  &          {}              & $2.20^{+0.33}_{-0.44}$    
\\\\[-1ex]
   28    &  mekal    &     0.68     &  15   & $1.38^{+0.63}_{-0.47}$  &
$1.95^{+1.45}_{-0.72}$  &          {}              & $1.09^{+0.50}_{-0.37}$    
\\\\[-1ex]\hline\\[-1ex]
   25    &  bb       &     1.43     &   8   & $0.0-0.13$              &
$1.12^{+0.27}_{-0.21}$  & $0.026_{-0.66}^{+0.81}$  & $0.340^{+0.91}_{-0.25}$ 
\\\\[-1ex]
   25    &  pl       &     0.77     &   8   & $0.0-0.3$               &
$0.82^{+0.43}_{-0.33}$  &                          & $0.425^{+0.35}_{-0.02}$ 
\\\\[-1ex]
   25    & mekal     &     1.11     &   8   & $0.17^{+0.31}_{-0.17}$  & $14.2 
-
79.80$          &                          & $0.376^{+0.60}_{-0.03}$ 
\\\\[-1ex] \hline
 \enddata

 \tablecomments{\\
\\
$^a$ bb =
blackbody; pl = power law; nsa = neutron star H-atmosphere;
mekal = optically-thin thermal plasma.\\
$^b$ The entry in this column depends on the spectral model --- it is the  
power law index $\Gamma$ or the temperature $kT$ in keV
($kT_{\rm eff}^\infty$ for the nsa model). \\
$^c$ Blackbody radius ($R_{\rm NS}^\infty$ for the nsa model).\\
$^d$ Unabsorbed flux in the 0.5--8.0 keV band.
}
 \end{deluxetable}

\end{document}